\documentclass[final,12pt,3p]{elsarticle}
\usepackage{graphicx,stfloats}
\usepackage{color}
\usepackage[colorlinks=true]{hyperref}
\usepackage{url}
\usepackage{blindtext}
\usepackage{latexsym,amsmath,amssymb}
\usepackage[T1]{fontenc}
\usepackage[utf8]{inputenc}
\usepackage[english]{babel}
\usepackage{csquotes}

\usepackage{algorithm2e}
\usepackage{scrextend}
\usepackage{mwe,tikz}
\usepackage[percent]{overpic}
\usepackage{color}
\usepackage[range-phrase={--},range-units=single]{siunitx}
\usepackage{graphicx}
\usepackage{tabularx}
\usepackage{subcaption}
\usepackage[export]{adjustbox}
\usepackage{wrapfig}
\usepackage{amsmath}
\graphicspath{{Figures/}}
\usepackage[version=4]{mhchem}
\usetikzlibrary{decorations.pathmorphing}
\tikzset{snake it/.style={decorate, decoration=snake}}

\usepackage[nameinlink, capitalize]{cleveref}
\newcommand{\ol}[1]{{\overline{#1}}}

\usepackage{booktabs,multirow,multicol}
\usepackage{todonotes}
\biboptions{sort&compress}

\journal{ }
\begin{document}
\begin{frontmatter}
\title{Interpretable Data-driven Methods for Subgrid-scale Closure in \\LES for Transcritical LOX/GCH4 Combustion}

\author[1]{Wai Tong Chung\corref{cor1}}
\ead{wtchung@stanford.edu}
\author[2]{Aashwin Ananda Mishra}
\author[1]{Matthias Ihme}

\address[1]{Department of Mechanical Engineering, Stanford University, Stanford, CA 94305, USA}
\address[2]{SLAC National Accelerator Laboratory, Menlo Park, CA 94025, USA}

\cortext[cor1]{Corresponding author:}

\begin{abstract}

Many practical combustion systems such as those in rockets, gas turbines, and internal combustion engines operate under high pressures that surpass the thermodynamic critical limit  of fuel-oxidizer mixtures. These conditions require the consideration of complex fluid behaviors that
pose challenges for numerical simulations, casting doubts on the validity of existing subgrid-scale (SGS) models in large-eddy simulations of these systems.
While data-driven methods have shown high accuracy as closure models in simulations of  turbulent flames, these models are often criticized for lack of physical interpretability, wherein they provide answers but no insight into their underlying rationale.
The objective of this study  is to assess SGS stress models from conventional physics-driven approaches and an interpretable machine learning algorithm, i.e., the random forest regressor, in a turbulent transcritical non-premixed flame. 
To this end, direct numerical simulations (DNS) of  transcritical liquid-oxygen/gaseous-methane (LOX/GCH4) inert and reacting flows are performed. 
Using this data, \textit{a priori}  analysis  is performed on the Favre-filtered DNS data to examine the accuracy of physics-based and random forest SGS-models under these conditions. 
SGS stresses calculated with the gradient model show good agreement with the exact terms extracted from  filtered DNS. 
The accuracy of the random-forest regressor decreased when physics-based constraints are applied to the feature set.
Results demonstrate that random forests can perform as effectively as algebraic models when modeling subgrid stresses, only when trained on a sufficiently representative database.  The employment of random forest feature importance score is shown to provide insight into discovering subgrid-scale stresses through sparse regression.

\end{abstract}


\end{frontmatter}

\section{Introduction}

The development of accurate computational tools has been essential for studying non-premixed flames within high pressure combustors.
Large-eddy simulations (LES) provide a feasible computational approach in capturing the behavior of flows within practical combustors. 
However, high pressure combustors in rockets and diesel engines operate under conditions that exceed the thermodynamic critical limits of both fuel and oxidizer.
Consequently, these conditions generate trans- and supercritical flows -- with complex behaviors that pose challenges for numerical modeling and simulations~\cite{Bellan2020}.
In many simulations, Lagrangian droplet methods are typically employed for simulating, which assumes the presence of liquid and gas phases. 
Simulations employing the Lagrangian droplet method have been shown to have good agreement with experimental measurements~\cite{CHANG1996,ZHU1996}.
However, these models often involve careful selection of the breakup and evaporation models along with parameter tuning.

Another approach for investigating high-pressure flows involves the use of the diffuse-interface method.
In contrast to sharp interface techniques, where interfaces are explicitly tracked or resolved in the computational domain, this method artificially diffuses the interface and treats the entire flow with a single real-fluid state equation.
While LES incorporating real-fluids effects have successfully been employed to simulate transcritical combustion \cite{Ma2019,Oefelein2019,Chung2020}, many of these simulations employ existing subgrid-scale  (SGS) models that have been typically developed for applications in subcritical pressure conditions~\cite{Vreman2004, Clark1979EvaluationFlow}.
 As a consequence, the application of these models to non-ideal flow regimes introduces uncertainties.
 One method for evaluating SGS models involves \emph{a priori} analysis, where  modeled SGS terms are compared with exact unclosed terms extracted from filtered DNS. 
 \citet{Selle2007} performed \textit{a priori} analysis on a three-dimensional DNS database  of supercritical binary mixtures   in turbulent mixing layers to demonstrate that the Smagorinsky model~\cite{Smagorinsky1963GENERALCE} performed poorly when predicting SGS stresses, while the gradient~\cite{Clark1979EvaluationFlow} and scale-similar~\cite{Bardina1980}  models performed well.
  In the same work, previously neglected unclosed terms for pressure and heat flux were shown to be  essential under supercritical conditions.
   \citet{Unnikrishnan2019} performed  \textit{a priori} analysis on two-dimensional DNS  of a transcritical reacting liquid-oxygen/gaseous-methane (LOX/GCH4) mixture to demonstrate that the mixed SGS model incorporating the dynamic Smagorinsky~\cite{Moin1991ATransport} was three times more accurate than the sole use of the dynamic Smagorinsky.

One approach for developing closure models in turbulent reacting flows involve the use of
data-driven methods.
 \emph{A priori} studies have been performed to demonstrate that convolutional neural-networks can provide accurate closure for turbulent combustion models~\cite{Lapeyre2019TrainingRates,Seltz2019DirectNetworks}.
 \citet{DEFRAHAN_SHASHANK_KING_DAY_GROUT_CF2019} demonstrated that deep learning models can generate as accurate results as  random forests, another regression model, with 25 fold improvement in computational costs when predicting the sub-filter probability density function in another \emph{a priori} study.
 \citet{Ranade2019AValidation} conducted an \emph{a posteriori} study to show that deep learning models can be trained with experimental data to  generate closure models for chemical scalars in Reynolds-averaged Navier–Stokes (RANS) simulations of turbulent jet flames. These studies have demonstrated the ability of deep learning  models in predicting closure models, but tend to offer little physical understanding. 
 
 Computational studies of high-pressure non-premixed flames were pioneered by Bray~et.~al. \cite{CHANG1996,ZHU1996,LAKSHMISHA1995}. 
 One work~\cite{LAKSHMISHA1995} examined the effect of different Damk\"{o}hler numbers (Da) on autoignition in high-pressure non-premixed flames under decaying homogeneous isotropic turbulence.
 In the present work, we perform DNS calculations of inert and reacting LOX/GCH4 non-premixed mixtures in the presence of decaying turbulence, under different Da, in order to evaluate algebraic and data-driven models for predicting unclosed SGS stresses under high pressure.
 Within this context, this \emph{a priori} study has the following objectives:
 \begin{itemize}
     \item To  identify and quantify limitations of  conventional algebraic SGS models in transcritical flows.
     \item To evaluate the insight gained from the random forest, an interpretable machine learning algorithm, in constructing a data-informed, interpretable SGS model.
 \end{itemize}
 The mathematical models for simulating the turbulent transcritical flows are presented in \cref{sec:method}. Details regarding the DNS configuration  are discussed in \cref{sec:dns}. \cref{sec:sgs} describes the SGS models and data-driven methods employed in the present work. Results from this \emph{a priori} study are discussed in \cref{sec:results}, before offering concluding remarks in \cref{sec:conc}.


\section{Mathematical Models} \label{sec:method}
The governing equations that are solved in the present study are the conservation equations for mass, momentum, energy, and chemical species:
\begin{subequations}
\begin{align}
    \partial_t {\rho} + \nabla \cdot ({\rho} {\boldsymbol{u}} ) &= 0\\
    \partial_t ({\rho} {\boldsymbol{u}}) + \nabla \cdot ({\rho} {\boldsymbol{u}} {\boldsymbol{u}} ) &= 
    - \nabla \cdot ({p}\boldsymbol{I})
      + \nabla \cdot {\boldsymbol{\tau}}_{v}\\ 
    \partial_t ({\rho} {e_t}) + \nabla \cdot [{\boldsymbol{u}} ({\rho}  {e_t} + {p})] &=  - \nabla \cdot {\boldsymbol{q}}_{v} +
    \nabla \cdot [ 
    ({\boldsymbol{\tau}}_{v}  )\cdot {\boldsymbol{u}}] 
    \\
    \partial_t ({\rho} Y_k) + \nabla \cdot ({\rho} {\boldsymbol{u}} Y_k) &= 
    -\nabla \cdot \boldsymbol{{j}}_{v} + {\dot{\omega}}_k \quad \textrm{where} \quad k = 1,2,...,N_s-1
\end{align}
\end{subequations}
with density $\rho$, velocity vector $\boldsymbol{u}$, pressure $p$,  specific total energy $e_t$, stress tensor $\boldsymbol{\tau}$, and  heat flux $\boldsymbol{q}$. $Y_k$, $\boldsymbol{j}_k$, and $ {\dot{\omega}_k}$ are the mass fraction, diffusion flux, and source term for the $k^{\textrm{th}}$ species, while subscript $v$ denotes viscous quantities. 

In the  \textit{a priori} analysis carried out in this study, a top-hat filter $H$ with a desired filter size $\overline{\Delta}$ is applied on a arbitrary quantity $\phi$ from the DNS data through a volume integral:
\begin{subequations}
\begin{align}
    \overline{\phi(\boldsymbol{x})} &= \int_V \phi(\boldsymbol{x}) H(\boldsymbol{x} - \boldsymbol{y},\ol{\Delta}) d\boldsymbol{y} \\
    H(\boldsymbol{x} - \boldsymbol{y},\ol{\Delta}) &= 
    \begin{cases}
        \frac{1}{\overline{\Delta}} &  \textrm{for }|\boldsymbol{x} - \boldsymbol{y}| \leq \overline{\Delta}/2, \\
        0 & \text{otherwise. }\\
    \end{cases}
\end{align}
\end{subequations}
and Favre-averaged quantity:
\begin{equation}
    \widetilde{\phi} = \frac{\overline{\rho \phi}}{\overline{\rho}}
\end{equation}
where $\ol{\,\cdot\,}$ denotes a filtered quantity and $\widetilde{\,\cdot\,}$ is a Favre-filtered quantity. 
After filtering, the governing equations become:
\begin{subequations}
\label{EQ_GOVERNING}
\begin{align}
    \partial_t \ol{\rho} + \nabla \cdot (\ol{\rho} \widetilde{\boldsymbol{u}} ) &= 0\\
    \partial_t (\ol{\rho} \widetilde{\boldsymbol{u}}) + \nabla \cdot (\ol{\rho} \widetilde{\boldsymbol{u}} \widetilde{\boldsymbol{u}} ) &= 
    - \nabla \cdot (\ol{p}\boldsymbol{I}) 
    + \nabla \cdot( \ol{\boldsymbol{\tau}}_{v}+  {\boldsymbol{\tau}}^{sgs})\\ \label{eq:LES}
    \partial_t (\ol{\rho} \widetilde{e}_t) + \nabla \cdot [\widetilde{\boldsymbol{u}} (\ol{\rho}  \widetilde{e}_t + \ol{p})] &= 
    - \nabla \cdot( \ol{\boldsymbol{q}}_{v}+ {\boldsymbol{q}}^{sgs})+\nabla \cdot [ 
    (\ol{\boldsymbol{\tau}}_{v} + {\boldsymbol{\tau}}^{sgs} )\cdot \widetilde{\boldsymbol{u}}] \\
    \partial_t (\ol{\rho} \widetilde{{Y}}_k) + \nabla \cdot (\ol{\rho} \widetilde{\boldsymbol{u}} \widetilde{{Y}}_k) &= 
    -\nabla \cdot( \boldsymbol{\ol{j}}_{v} + \boldsymbol{{j}}^{sgs})+ \ol{\dot{{\omega}}}_k \quad \textrm{where} \quad k = 1,2,...,N_s-1 \label{eq:speciesTrans}
\end{align}
\end{subequations}
with superscript  $sgs$ denoting  subgrid-scale quantities. Exact subgrid-scale quantities can then be extracted directly from the filtered DNS or approximated through the models described in \cref{sec:sgs}.


The Peng-Robinson cubic state equation~\cite{Peng76} is employed to model real-fluid thermodynamics under transcritical conditions:
\begin{equation}
\label{eq:pr}
    p = \frac{\rho RT}{1-b\rho}-\frac{a\rho^2}{1 + 2b\rho -b^2\rho^2}
\end{equation}
with mixture-specific gas constant $R$. The coefficients $a$ and $b$ account for effects of intermolecular forces and volumetric displacement, and are dependent on temperature and composition \cite{Poling2001}. 
Details regarding the evaluation of specific heat capacity, internal energy, and  partial enthalpy from the Peng-Robinson state equation is described in \citet{Ma2017}

In this study, the two-step five-species CH4-BFER mechanism~\cite{FRANZELLI2012621} is employed, which has been applied to investigate a supercritical gas-turbine combustor at 20 MPa~\cite{Chong2017}. 
In DNS of trans- and supercritical combustion,  reduced chemical mechanisms~\cite{BELLAN2017245,BUSHE2020131} have been employed to circumvent large computational costs incurred by solving non-ideal state equations. 
Takahashi's high-pressure correction~\cite{Takahashi1975} is used to evaluate the binary diffusion coefficients.
Since only two species are used in the inert simulations, the binary diffusion coefficients are exact. Thermal conductivity  and dynamic viscosity are evaluated using Chung's method with high-pressure correction~\cite{Chung1988}. 
For multi-species mixtures in the reacting cases,  Chung's pressure correction is known to produce oscillations, especially for dynamic viscosity.
Hence, transport properties of the mixture are evaluated through  mole-fraction-averaging of individual species.
A similar approach has been applied in prior studies~\cite{Oefelein2019,Unnikrishnan2019}.

Simulations are performed by employing an unstructured compressible finite-volume solver~\cite{Khalighi_2011,Ma2017,chung2021}. A central scheme, which is 4th-order accurate on uniform meshes, is used along with a 2nd-order ENO scheme. 
The ENO scheme is activated only in regions of high local density variations using a threshold-based sensor to describe sharp interfaces present in transcritical flows. 
Due to the density gradients present at trans- and supercritical conditions, an entropy-stable flux correction technique~\cite{Ma2017} is used  to dampen non-linear instabilities in the numerical scheme. 
 The double-flux method by Ma et al. \cite{Ma2017} is used with a dynamic sensor to eliminate spurious pressure oscillations.
A Strang-splitting scheme is employed for time-advancement, combining a strong stability preserving 3rd-order Runge-Kutta (SSP-RK3) scheme for integrating the non-stiff operators with a semi-implicit scheme~\cite{Wu2019EfficientChemistry} for advancing the chemical source terms. 


\section{DNS Configuration }
\label{sec:dns}
\begin{figure*}[h]
    \centering
    \includegraphics[width=\textwidth]{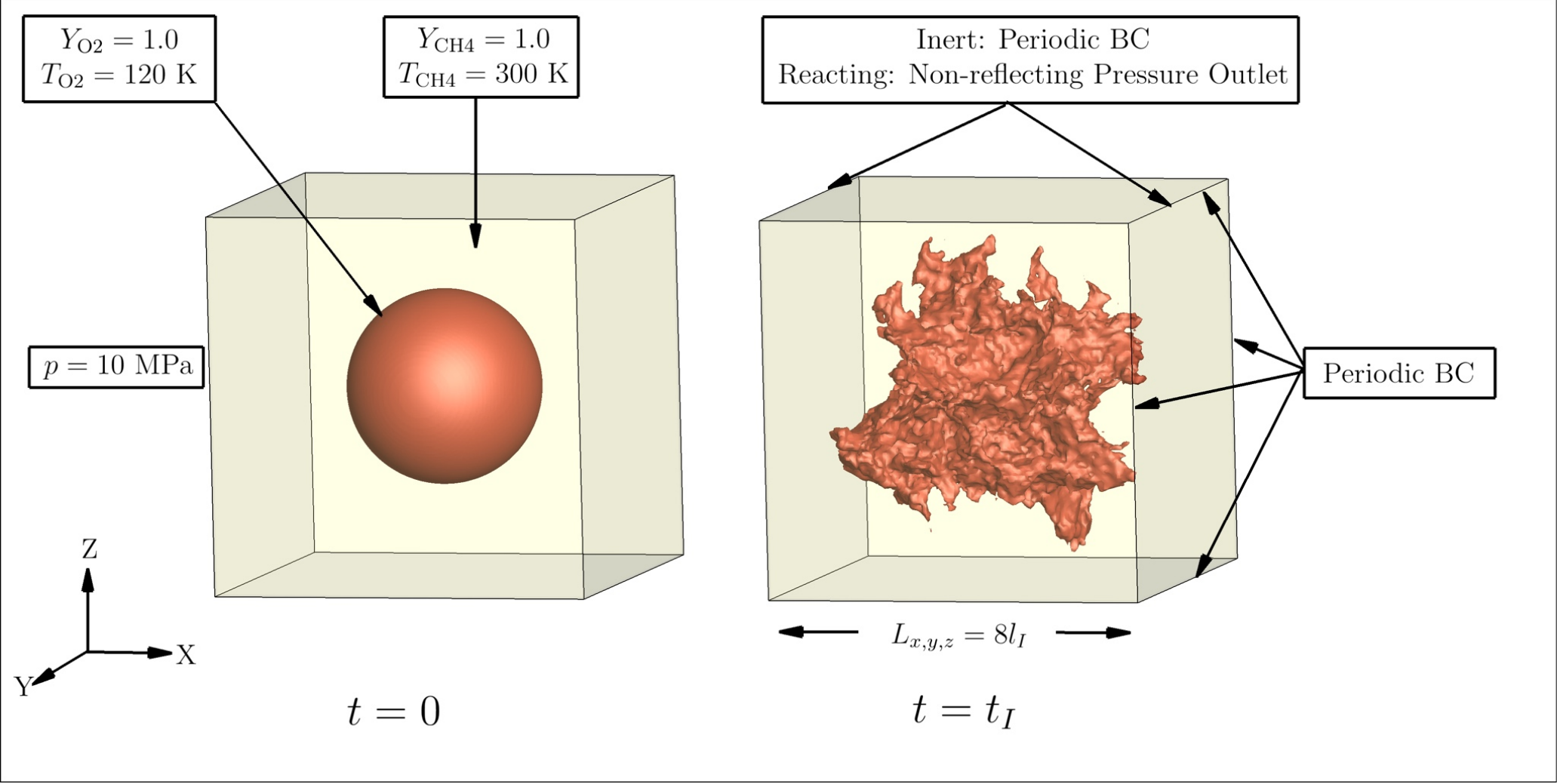}
    \caption{DNS investigated at initial time $t = 0$ and one eddy turnover time $t = t_I$. Isosurface shows stoichiometric mixture fraction $Z=0.2$ for the inert case.}
    \label{fig:setup}
\end{figure*}

Inert and reacting direct numerical simulations are performed on  a three-dimensional cubic domain, with length $L$, of  LOX/GCH4 mixture  shown in \cref{fig:setup}.
In this setup, a spherical liquid oxygen core, with a radius $r=0.25L$, is initialized in gaseous methane, where the radial profile of the initial conditions is chosen to match one-dimensional calculations of inert and reacting mixtures under the same conditions.
Fuel and oxidizer temperatures, $T_{\textrm{CH4}}$ and $T_{\textrm{O2}}$ have been set at 120 K and 300 K, respectively, while the pressure is set at 10 MPa.
These operating conditions are chosen to match practical LOX/GCH4 combustors, and were investigated in previous studies~\cite{Huo2011,Unnikrishnan2018B}. Periodic boundary conditions are used for all boundaries in  the inert case. In the reacting cases, non-reflecting pressure outlets are used in both boundaries in the $x$-direction, while remaining boundaries are periodic.

The initial velocity profile was generated with a synthetic isotropic turbulence generator by \citet{Saad2017ScalableSpectra} with zero mean velocity, based on the von K\'arm\'an-Pao energy spectrum:
\begin{subequations}
\begin{align}
    E(\kappa) &= \alpha \frac{u'^2}{\kappa_I}\frac{(\kappa/\kappa_I)^4}{[1+(\kappa/\kappa_I]^{17/6}}\exp \left[- 2 \left( \frac{\kappa}{\kappa_\eta}\right)^2\right], \\ 
    \alpha &= 1.453, \\
    \kappa_I &= 0.746834/l_I,
\end{align}
\end{subequations}
where $u'$ is the RMS velocity, $\kappa$ is the wave number, and $\kappa_\eta$ the Kolmogorov wave number. The chosen scaling constant $\alpha$ and large-eddy wavenumber  $\kappa_I$ are typical for isotropic turbulence \cite{Bailly1999AEquations}.
In all cases, the integral lengthscale $l_I$ and root-mean-squared (RMS) velocity fluctuation $u'$ have been chosen to produce a turbulent Reynolds number Re$_t$ of 80, which has been computed with the averaged kinematic viscosity of oxygen and methane  at 120 K and 300 K, respectively. 

\begin{figure}[htb!]
        \centering
    \begin{subfigure}{0.32\columnwidth}
        \centering            \includegraphics[width=\columnwidth]{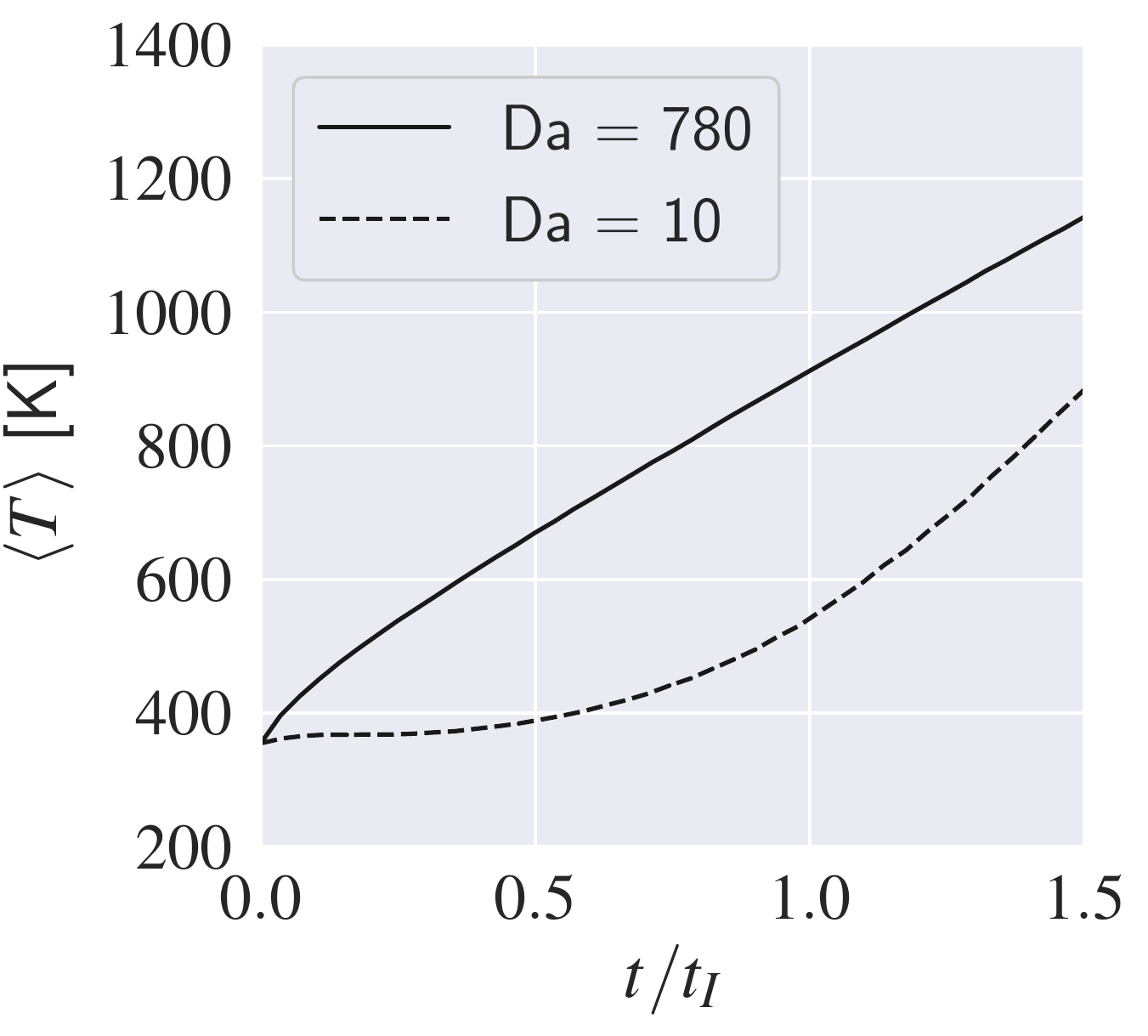}
        \caption{Temperature.\label{fig:DaTemp}}
    \end{subfigure}
        \begin{subfigure}{0.3\columnwidth}
        \centering            \includegraphics[width=\columnwidth]{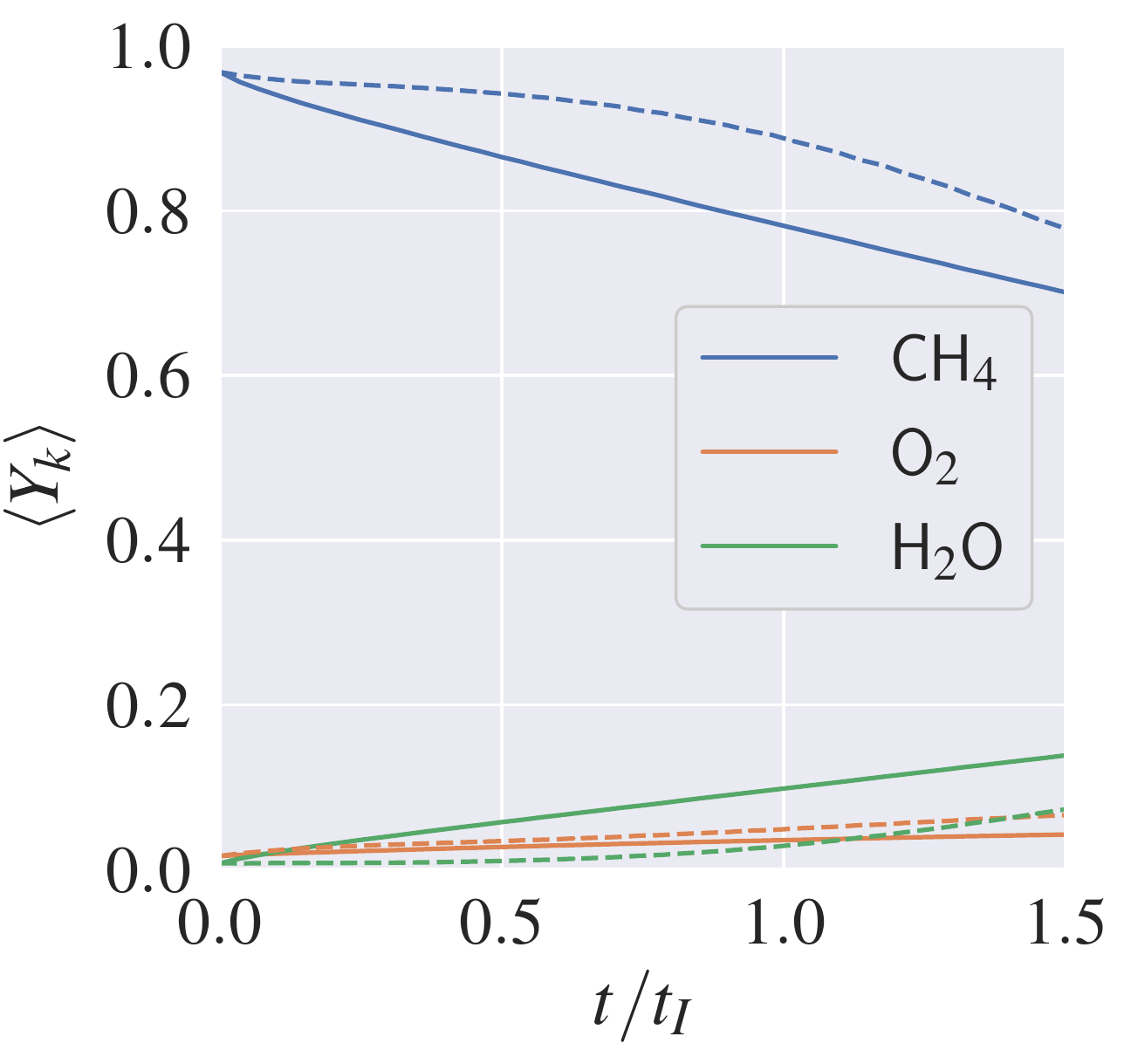}
        \caption{Mass fraction.\label{fig:DaSpecies}}
    \end{subfigure}
    \begin{subfigure}{0.3\columnwidth}
        \centering
            \includegraphics[width=\columnwidth]{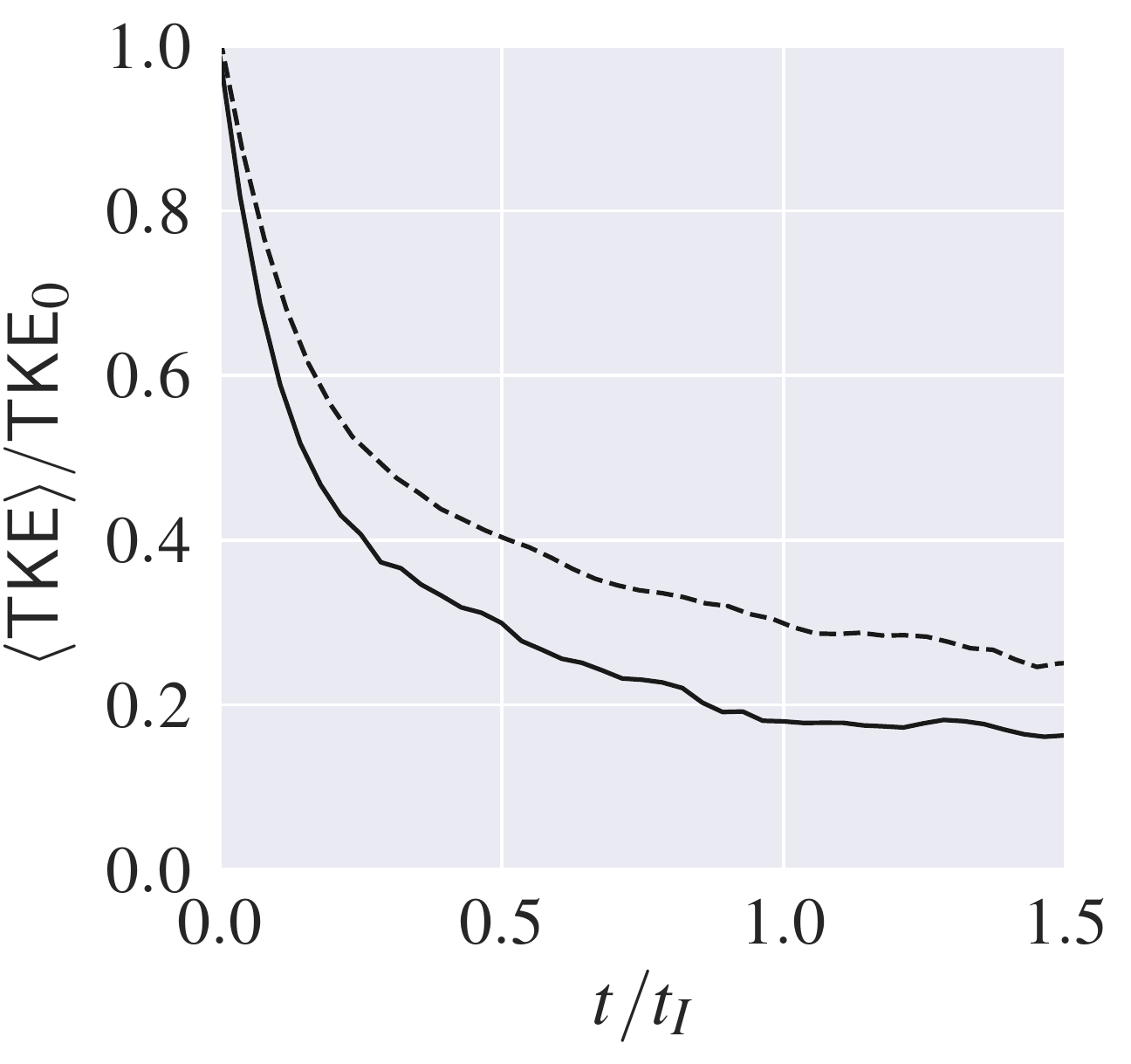}
        \caption{Turbulent kinetic energy. \label{fig:DaTKE}}
    \end{subfigure}
    \caption{ Temporal evolution of  global temperature $T$, mass fraction $Y_k$, and normalized turbulent kinetic energy TKE for two reacting cases.\label{fig:Datime}.}
\end{figure}

In the reacting cases, two different Damk\"ohler numbers, Da, of 780 and 10 are investigated, corresponding to flamelet and unsteady regimes~\cite{Williams2006DescriptionsCombustion}, respectively. 
The Damk\"ohler number is given by the ratio of physical timescale $t_{phys}$ and chemical timescale $t_{chem}$:
\begin{equation}
    \textrm{Da} = \frac{t_{phys}}{t_{chem}},
\end{equation}
where  $t_{chem}$ is approximated from the extinction strain rate of a one-dimensional counterflow diffusion flame of a LOX/GCH4 mixture under similar conditions, and physical time is evaluated from the eddy turnover time $t_{phys}=t_I$.
\Cref{fig:DaTemp} shows that the mean temperature $\langle T \rangle$ is lower when Da~=~10 than when Da~=~780, due the presence of local extinction. This is also reflected in \Cref{fig:DaSpecies} where the consumption of CH$_4$ is slower in the case Da~=~10 than the case Da~=~780. 
This decrease in temperature and composition also results in a slower decay of turbulence, as shown by the mean turbulent kinetic energy $\langle \textrm{TKE} \rangle$, normalized by the initial TKE, shown in \cref{fig:DaTKE}.

\begin{figure}[htb!]
        \centering
    \begin{subfigure}{\columnwidth}
        \centering
            \includegraphics[width=0.33\columnwidth]{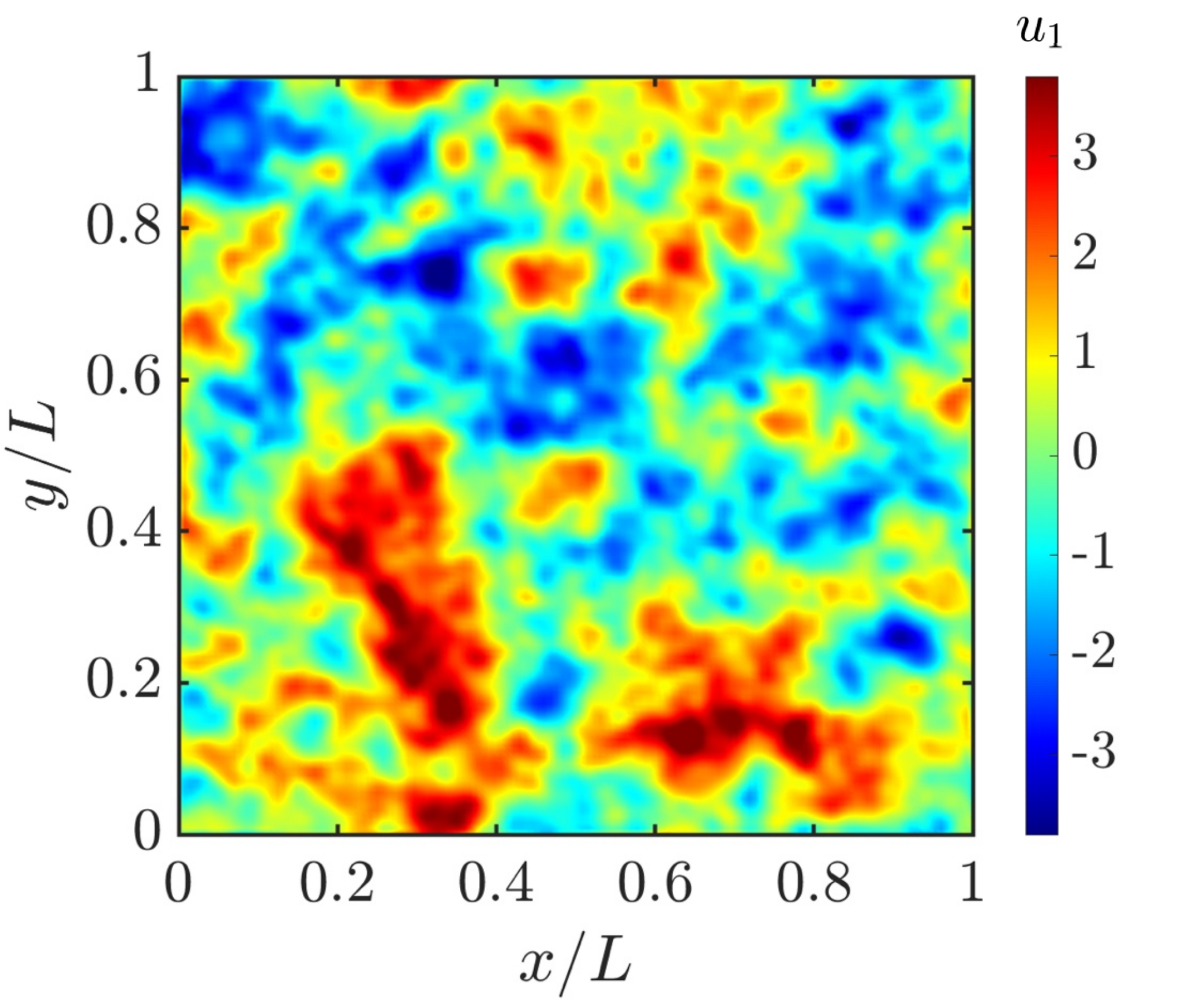}
            \includegraphics[width=0.33\columnwidth]{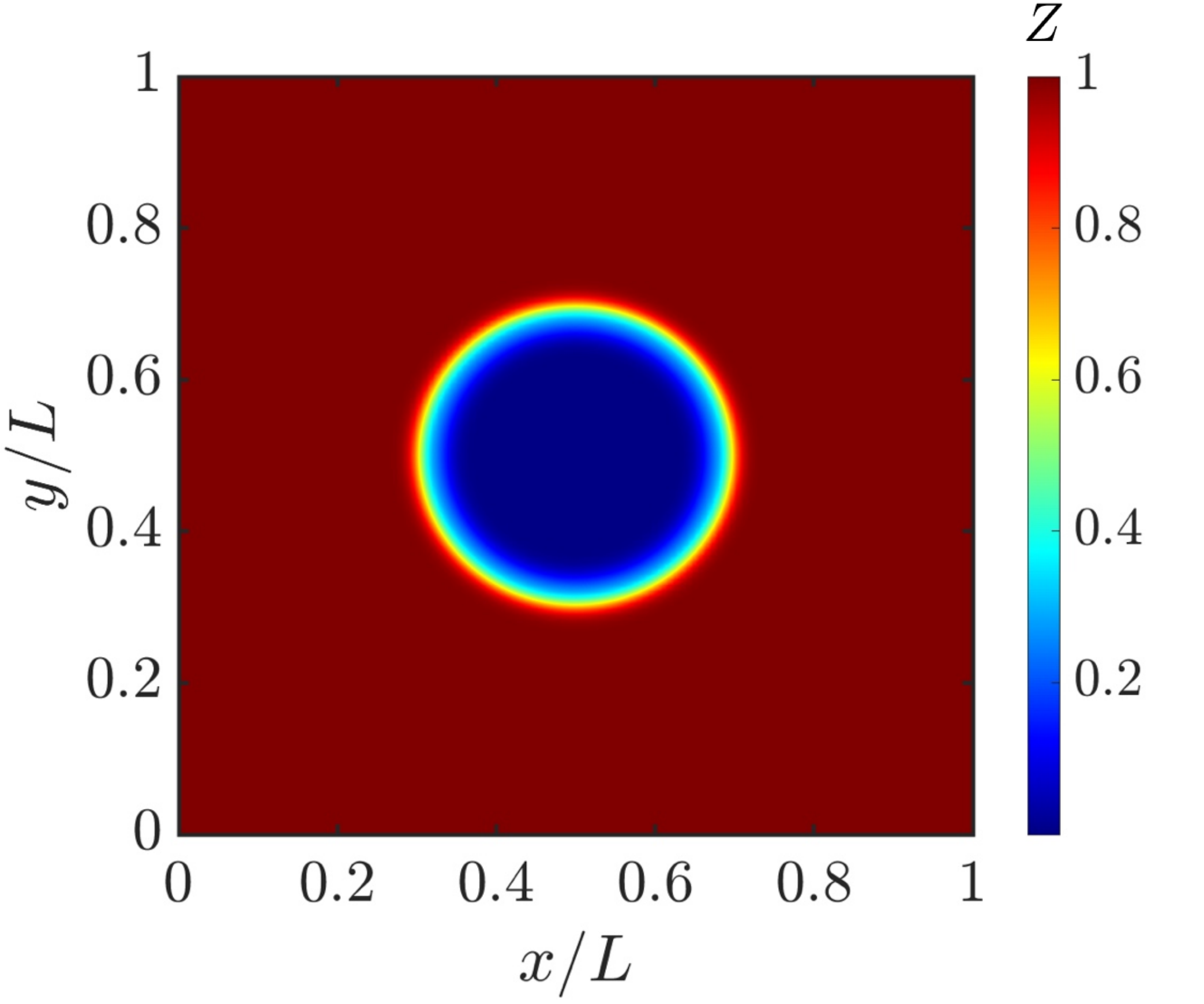}
            \includegraphics[width=0.3\columnwidth]{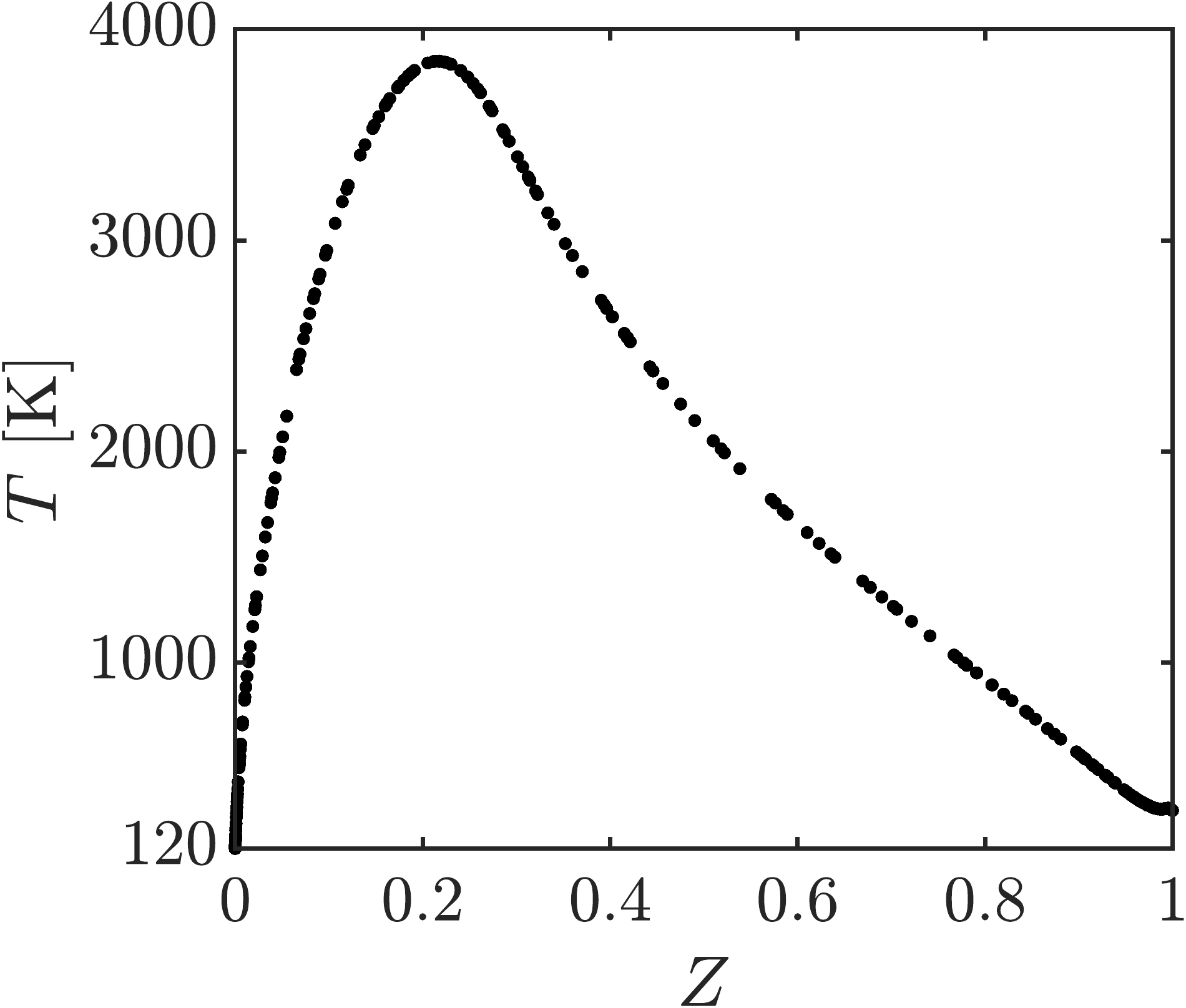}
        \caption{Case Da = 10 at initial time $t = 0$.}
    \end{subfigure}

    \begin{subfigure}{\columnwidth}
        \centering
            \includegraphics[width=0.33\columnwidth]{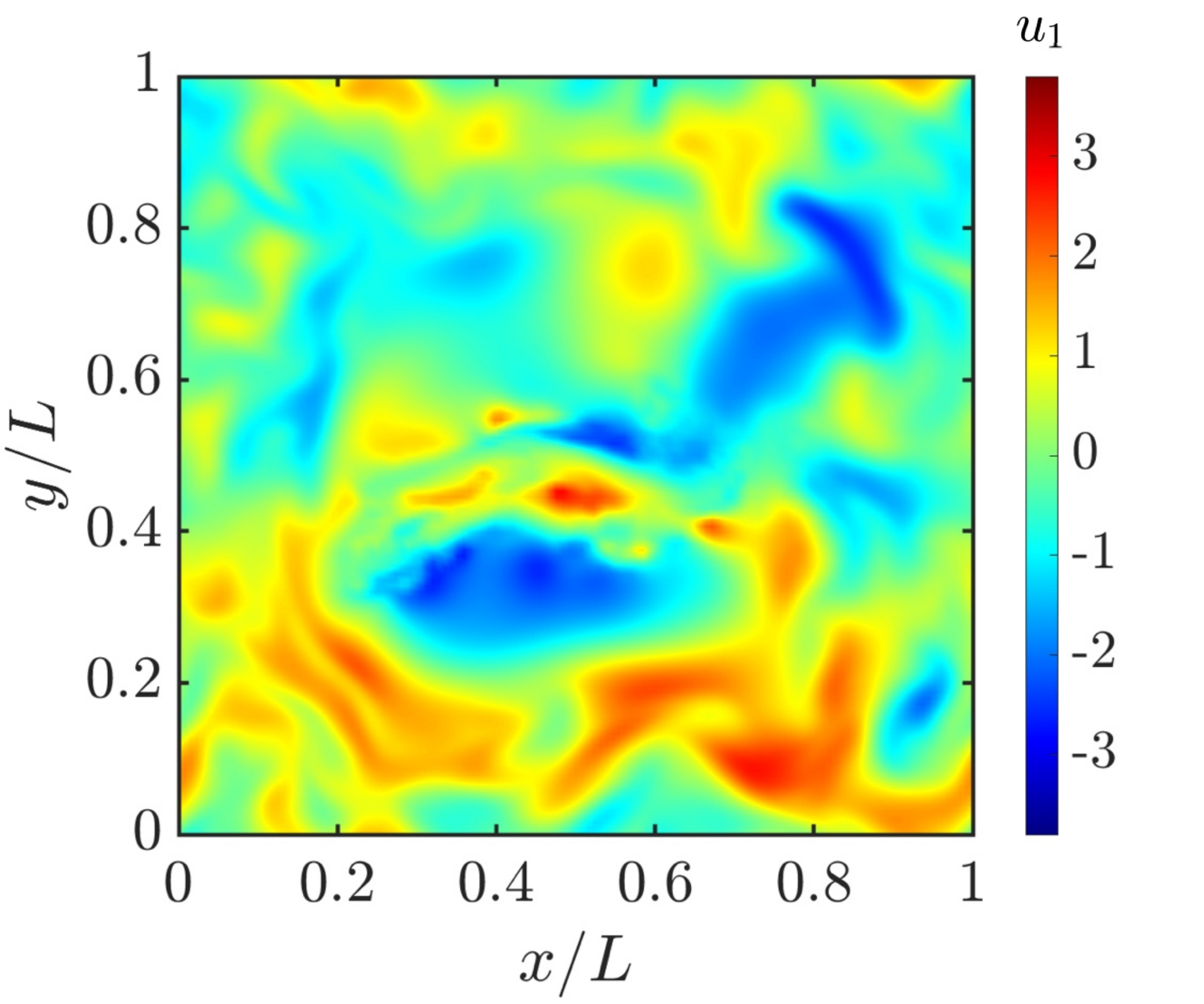}
            \includegraphics[width=0.33\columnwidth]{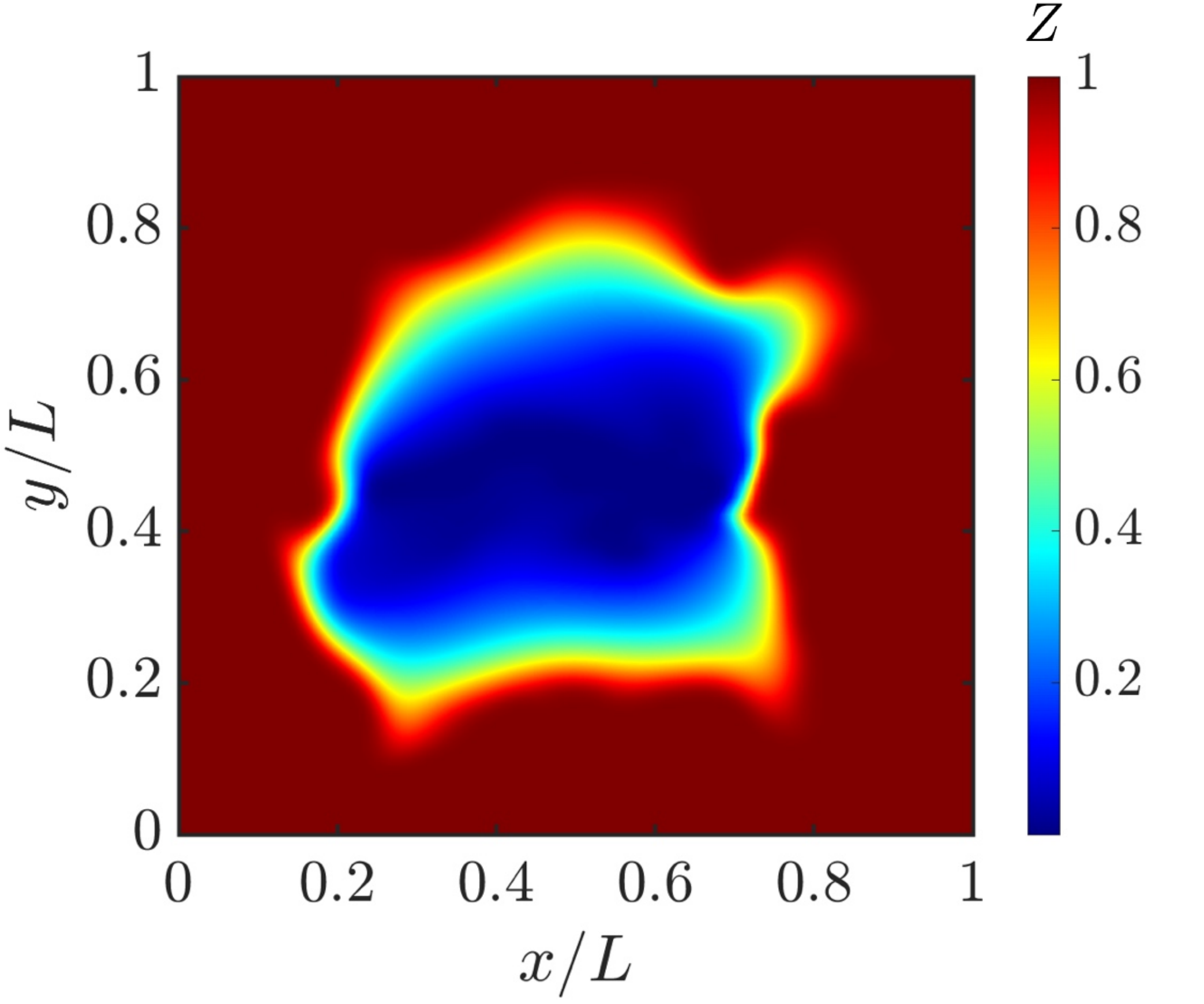}
            \includegraphics[width=0.3\columnwidth]{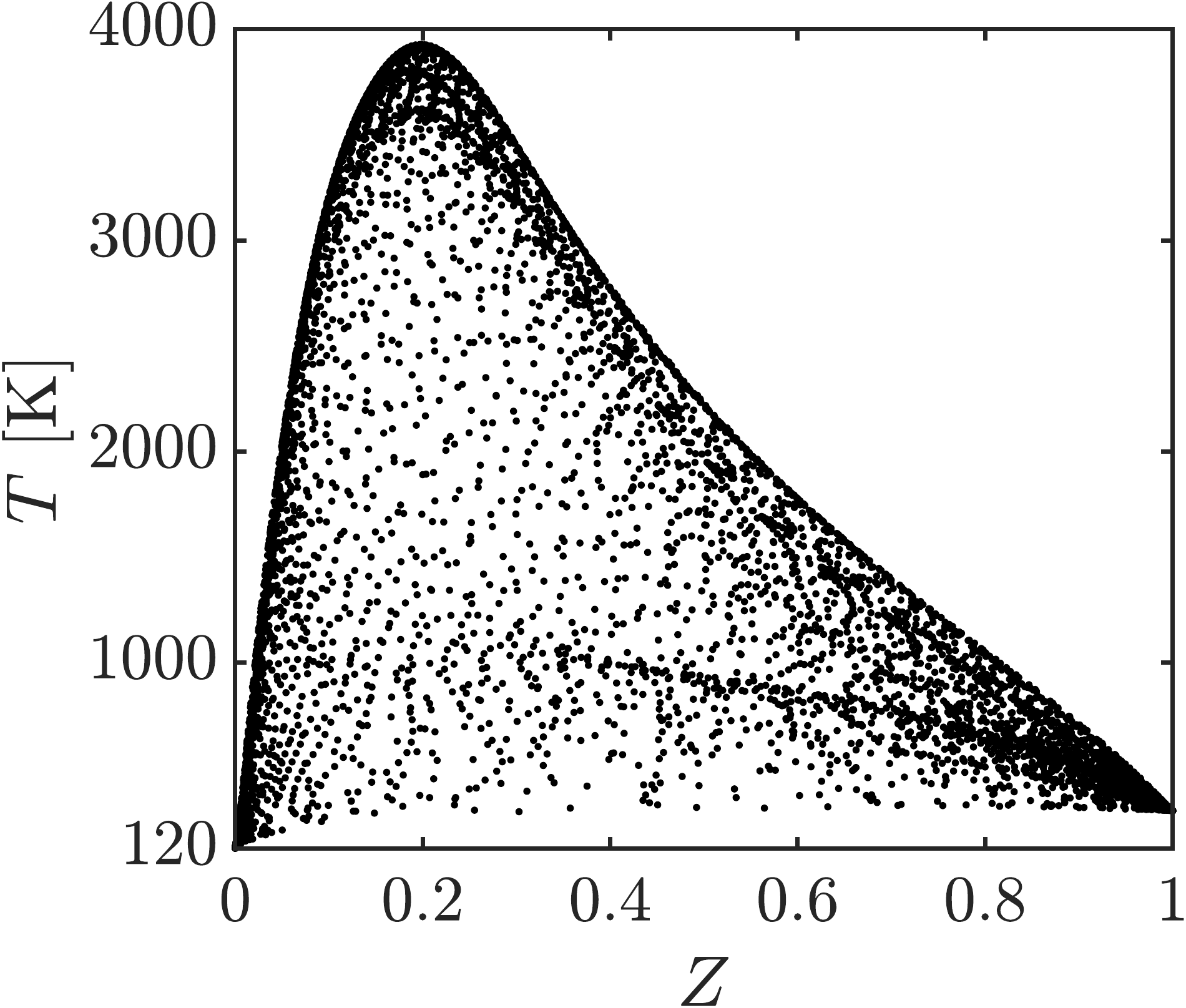}
        \caption{Case Da = 10 at eddy turnover time $t = t_I$.}
    \end{subfigure}
        \begin{subfigure}{\columnwidth}
        \centering
            \includegraphics[width=0.33\columnwidth]{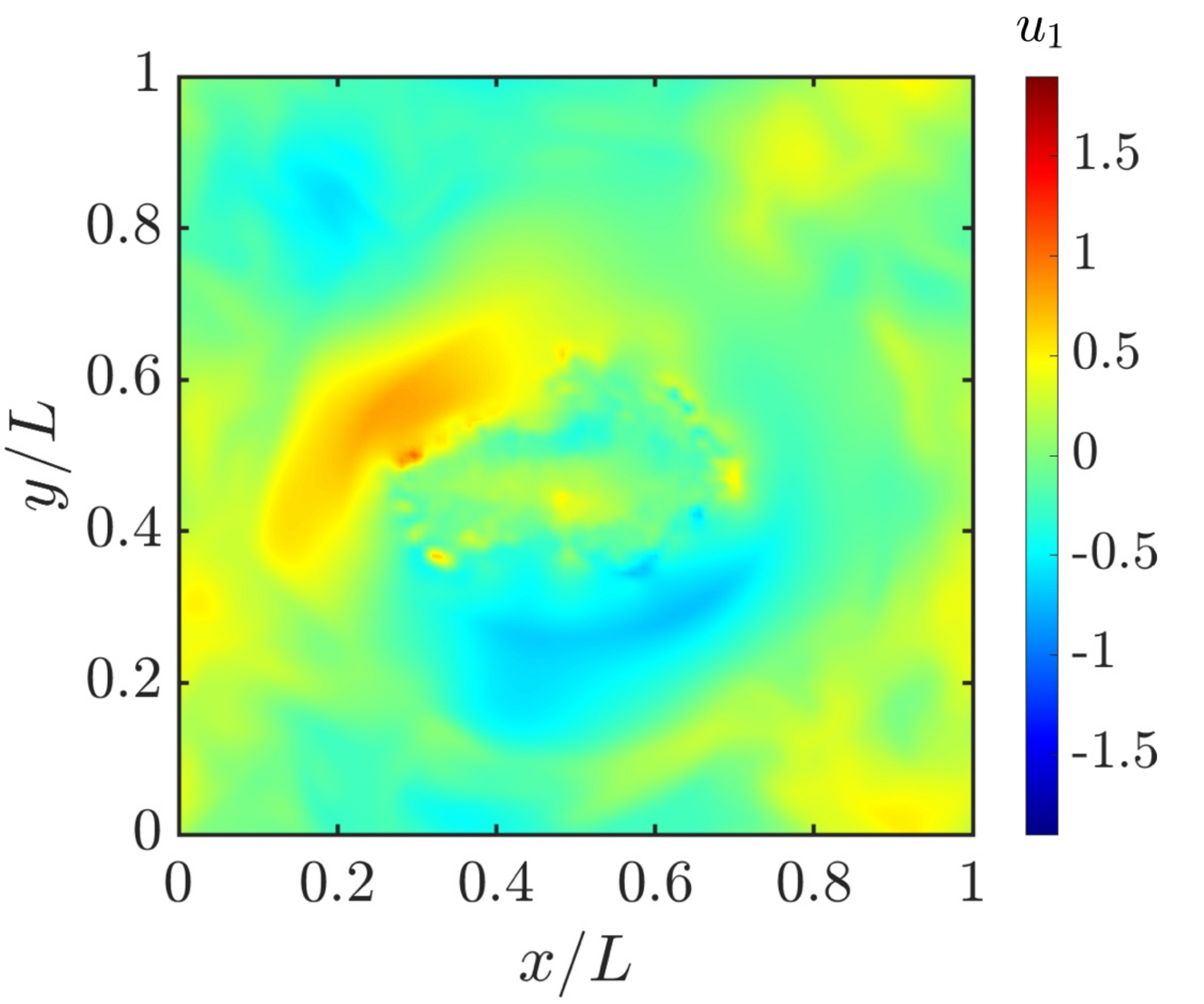}
            \includegraphics[width=0.33\columnwidth]{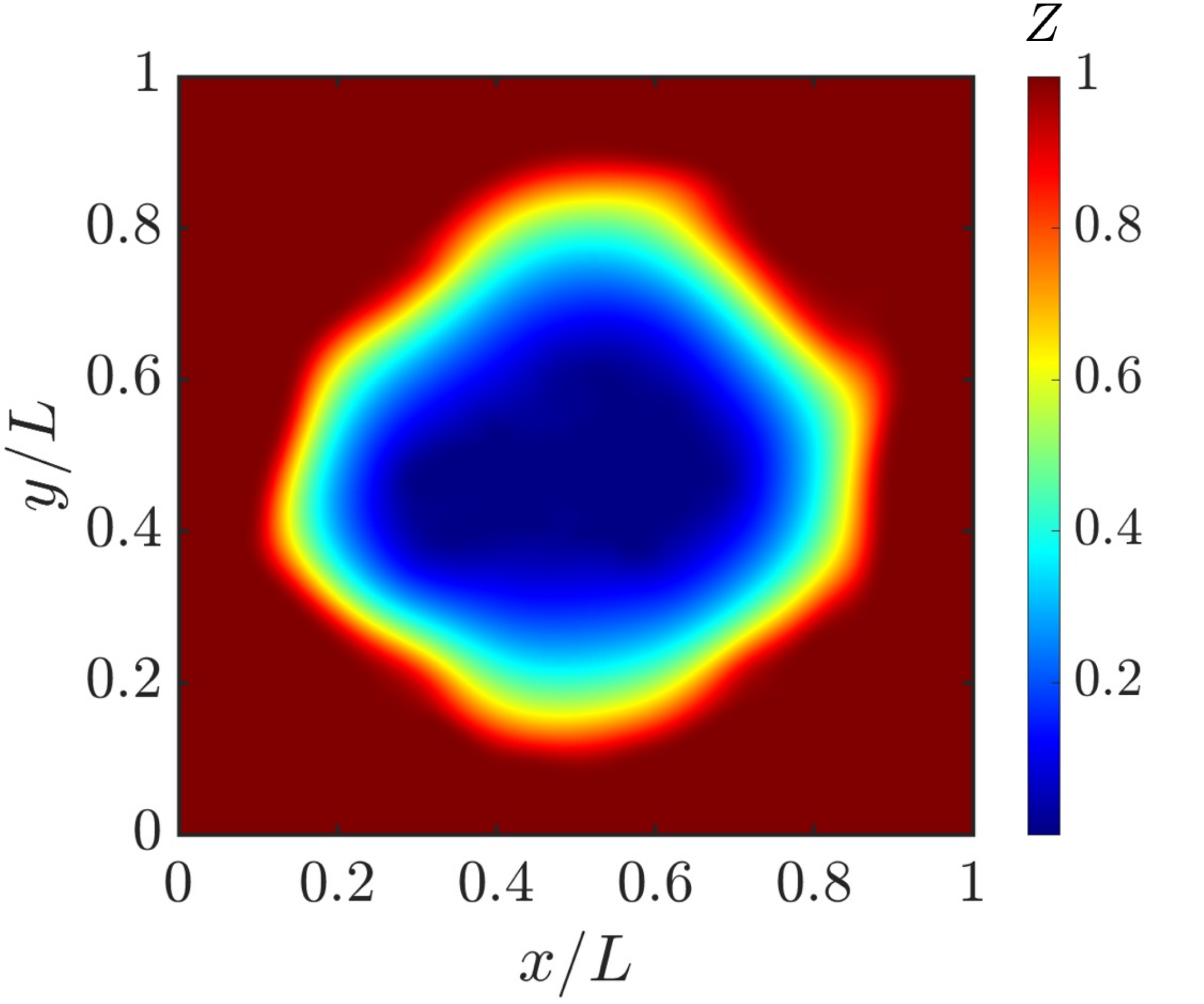}
            \includegraphics[width=0.3\columnwidth]{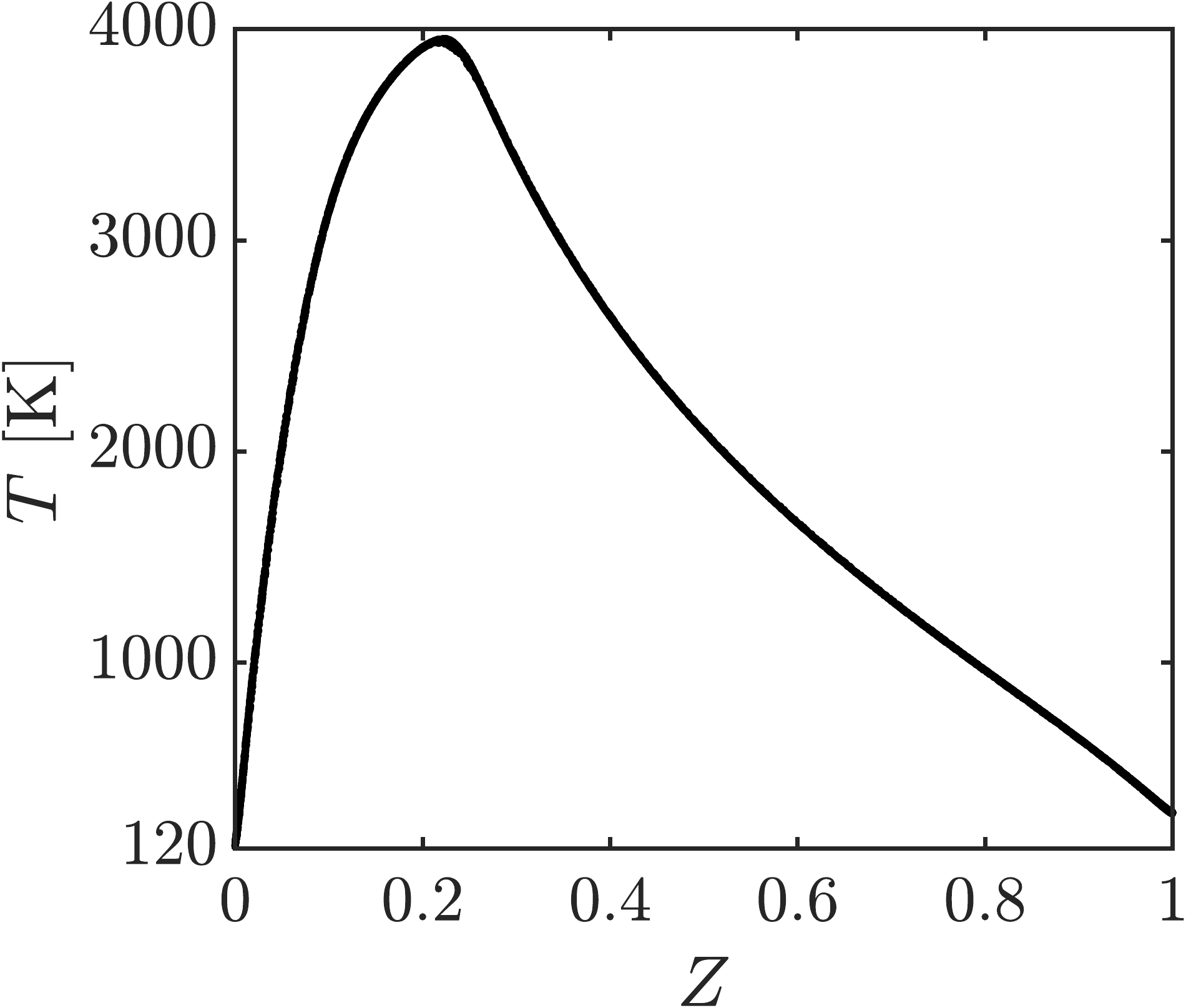}
        \caption{Case Da = 780 at eddy turnover time $t = t_I$.}
    \end{subfigure}
    \caption{ Axial velocity $u_1$, mixture fraction $Z$, and conditional temperature $T$ for the reacting cases at transverse location $z=0$.\label{fig:flowfield}}
\end{figure}

In this study, analysis is performed on all cases after $t = \textrm{argmax}(t_I,t_{chem})$, which is typically done for DNS of combustion under decaying turbulence in order to ensure  the flow fields are independent of initialization~\cite{SCHOEPPLEIN20181166}. Instantaneous flow fields for axial velocity component $u_1$, mixture fraction $Z$, and mixture-fraction conditioned temperature $T$ for the reacting cases at $t=  0$ and $t = t_I$ are shown in \cref{fig:flowfield}.

\Cref{tab:init} summarizes the DNS cases examined in this study.
The domain lengths in all direction were chosen to be eight times the size of the integral lengthscale $l_I$ to minimize effects of the boundary conditions. Cell size $\Delta$ is prescribed in the order of the Kolmogorov lengthscale, ensuring that all lengthscales are resolved. In addition, a mesh refinement study was performed, where the energy spectra of velocity was found to converge between $128^3$ and $256^3$.
Simulations for all three cases are advanced with a convective CFL number of unity, corresponding to timesteps of 2.5 and 0.5 ns for cases Da = 780 and Da = 10, respectively. The simulations were performed using 960 Intel Xeon (E5-2698 v3) processors, and 2.3 $\mu$s and 0.6 $\mu$s of physical time could be completed in about an hour wall clock time for cases Da = 780 and Da = 10, respectively. 

\begin{table}[ht]
 \centering
 \caption{Summary of DNS cases. \label{tab:init}}
 \resizebox{0.7\columnwidth}{!}{
\begin{tabular}{|l|ccccccc|}
\hline
Case
&$N_{x,y,z}$
& $L_{x,y,z}$ [$\mu$m]
	
&  Re$_t$	
& 	$l_I$ [$\mu$m]	
& 	$\eta_k$ [$\mu$m]
& 	$\Delta$ [$\mu$m]
& $t_I$ [$\mu$s]\\
\hline\hline
Inert
&$128$
& 500
& 	80
&  62.5	
& 2.32	
& 3.91	& 286\\
Da = 780
&$128$
& 500
& 	80
& 	62.5	
& 	2.32	
& 	3.91 &286	\\
Da = 10
&$128$
& 60
& 	80
& 7.50	
& 0.278		
& 0.469	& 4.12\\
\hline
\end{tabular}
}
\end{table}

\section{Subgrid-scale models and data-driven methods}
\label{sec:sgs}
In this study, we employ conventional algebraic and data-driven methods for predicting the subgrid-scale fluxes from the LES momentum equation (\cref{eq:LES}):
\begin{equation}
    \tau_{ij}^{sgs} = \ol{\rho}(\widetilde{u_i u_j} - \widetilde{u}_i\widetilde{u}_j)
    \label{eq:sgs}
\end{equation}
Two algebraic SGS models, namely the Vreman and the gradient model,  and random forest regressors  are evaluated.
Additionally, we demonstrate the employment of random forest feature importance scores for assisting the discovery of algebraic SGS models by sparse regression. 

\subsection{Algebraic models}
The Vreman SGS model~\cite{Vreman2004} is derived from  the eddy-viscosity hypothesis:
\begin{equation}
    \tau_{ij}^{sgs} \simeq -2\ol{\rho}\nu_{\text{SGS}}\widetilde{S}_{ij} + \frac{1}{3}\tau_{kk}\delta_{ij},
    \label{eq:vreman}
\end{equation}
where  $S_{ij}$ is the velocity strain tensor, and $\delta_{ij}$ is the Kronecker delta. The eddy viscosity $\nu_{\text{SGS}}$ is evaluated for a filter width $\ol{\Delta}$ as follows:
\begin{subequations}
\begin{align}
    \nu_{\text{SGS}} &= C_v \sqrt{\frac{B}{a_{ij}a_{ij}}},\\
    a_{ij} &= \dfrac{\partial \widetilde{u}_i}{\partial x_j},\\
    B &= \beta_{11}\beta_{22} - \beta_{12}^2 + \beta_{11}\beta_{33} - \beta_{13}^2 + \beta_{22}\beta_{33} - \beta_{23}^2,\\
    \beta_{ij} &= \ol{\Delta}^2a_{ki}a_{kj},
\end{align}
\end{subequations}
where a Vreman coefficient $C_v$ of 0.07 is typically  used in isotropic turbulence.

The gradient model by \citet{Clark1979EvaluationFlow} is extracted from the first term in the Taylor series expansion of the filtering operation, and is given by:
\begin{equation}
    \tau_{ij}^{sgs} \approx \ol{\rho}C_{g}\ol{\Delta}^2\dfrac{\partial \widetilde{u}_i}{\partial x_k}\dfrac{\partial \widetilde{u}_j}{\partial x_k},
    \label{eq:gradient}
\end{equation}
where a coefficient $C_{g}$ of 1/12 is typically used when a top-hat filter is employed.  In the present study, we wil evaluate both models and compare results against DNS data and a data-driven approach.

\subsection{Random forest regressor}
In this study, we employ the random forest as our regression algorithm.
Random forests~\cite{Breiman2001RandomForests} consist of an ensemble of decorrelated Classification and Regression Trees (CARTs) \cite{Breiman1984ClassificationTrees}. CARTs are a machine learning approach for formulating prediction models from data by recursively partitioning the inputted feature space, and fitting a simple prediction within each final partition. The partitioning of the feature space can be represented as a decision trees. Decision trees are supervised graph based model wherein the tree consists of nodes and edges. The internal (or non-terminal) nodes of the tree represent splits based on learned partitions of the feature space. Each leaf (or terminal) node is associated with a numerical value for regression trees (as opposed to categorical targets for classification trees). 

During the training phase, the structure of the decision tree and the partitions associated with each node are inferred. During each step of this phase, exhaustive sets of splits over different input features are evaluated. The split leading to maximal decrease in prediction variance is selected at the associated node. The procedure continues to make recursive splits based on dataset until it has reduced the overall variance below a given threshold or upon reaching a given stopping parameter (for instance, upon reaching a maximal depth of the tree).

During the prediction phase, a new sample is traversed down the tree from the root node to a leaf node, wherein its path is determined based on the partition at each internal node. Once a leaf node is reached, the numerical value associated with the specific leaf node is outputted as the prediction for the sample. 

Decision trees are non-parametric and can model arbitrarily complex relations without any \emph{a priori} assumptions, but  are prone to overfitting due to the greedy nature of the inference algorithm. Thus, decision trees have low bias but high variance. 

Random Forests are an ensemble learning algorithm, wherein the predictions of ensembles of decorrelated decision trees are aggregated so as to give a final meta-model with low bias and low variance. The decorrelation amongst the individual trees in the ensemble is achieved using bootstrapping~\cite{Breiman2001RandomForests} in conjunction with feature bagging~\cite{amit1997joint} during the training of each decision tree. The final prediction of the resulting ensemble model is by averaging the predictions of all trained individual trees (or equivalently, via aggregation).

While accurate predictions are an important goal for machine learning models, in many fields of application it is just as important to derive understanding from the model. At the basic level, this can be embodied via feature importances, wherein the trained model also provides information regarding importances of different input features to the final prediction. Such measures of model interpretability provide insight into the underlying rationale learned by the model during training and can lead to high confidence in the model. Similarly, such interpretability measures can lead to data-driven discovery of new relationships between the input features and targets. Random forests provide the Mean Decrease Impurity  (MDI) importance measure for all the features in the input set~\cite{louppe2013understanding}. Here, the importance of a feature is given by aggregating the weighted decrease in variance for all the nodes where the specific feature is used as the criterion for partitioning the feature space. 

In the present investigation, the random forest regressor implementation from the Scikit-learn library~\cite{scikit-learn} is used.
Here, a random forest consisting of fifty decision trees is employed. 
The hyperparameters of the random forest are determined using a random grid search approach with a 3-fold cross-validation set. 
Training is performed once \emph{a priori}, and requires 88s of walltime with 8 CPUs, when trained on data coarsened for three different filter sizes from a single timestep. Prediction time for a $64^3$ dataset requires 2.4 s on a single CPU.

\subsection{Sparse regression for model discovery}
\label{sec:sparse}
A linear model $\hat{f}_i$ for $m$ number of samples is typically expressed as the weighted sum of independent quantities $X$:
\begin{equation}
    \hat{f}_i = \sum_{j=1}^n \beta_j X_{ij} \quad 0 \leq i \leq m
    \label{eq:linearmodel}
\end{equation}
with $n$ number of model coefficients $\beta$ being employed. 

When samples of the ground truth \textbf{f} are available, the model coefficients can be found with the $l_1$-norm regularized least-squares or \emph{lasso} method~\cite{Tibshirani1996}:
\begin{align}
\min_{\boldsymbol{\beta}} \left\{ \frac{1}{m} ||\mathbf{f}-\mathbf{X}\boldsymbol{\beta} ||^2_2 + \lambda ||\boldsymbol{\beta}||_1 \right\} \label{eq:lsqr},
\end{align}
where $\lambda$ is a regularization parameter for controlling the tradeoff between the least-squares fit and the $l_1$-norm. 
Since, the optimization scheme in \cref{eq:lsqr}  also minimizes the $l_1$-norm of the model coefficients $||\boldsymbol{\beta}||_1$, lasso  encourages sparsity, i.e. reduces the number of terms in the linear model, as  zero-valued model coefficients are preferred.

In the context of discovering subgrid-scale models, non-linearities  can be introduced by replacing \textbf{X} non-linear functions $G(\mathbf{X})$ of the original variables. In this study, we construct a model with non-linear variables by evaluating $d$-order  polynomial functions:
\begin{subequations}
\begin{align}
\hat{\boldsymbol{f}}&=G^d (\mathbf{X}) \boldsymbol{\beta}, \\
    G^d (\mathbf{X}) &= 
\begin{bmatrix}
1 & X_{11} &  X_{12} &\cdots& X_{1n} & X_{11}^2 & X_{11} X_{12} & \cdots &  X_{1n}^d \\
\vdots & \vdots & \vdots & \vdots & \vdots & \vdots &\vdots & \vdots &\vdots   \\
1 & X_{m1} & X_{mn} &\cdots& X_{mn} & X_{m1}^2 & X_{m1} X_{m2} & \cdots &  X_{mn}^d 
\end{bmatrix}, \label{eq:matrix}\\
\boldsymbol{\beta} &= 
\begin{bmatrix}
\beta_0 & \beta_1 & \cdots & \beta_k 
\end{bmatrix}^T, \quad k = \sum^d_{i=1} n^i
\end{align}
\label{eq:nonlinearcoef}
\end{subequations}

\Cref{eq:nonlinearcoef} demonstrates that the dimensionality of this approach scales to order of polynomial functions $\mathcal{O}(mn^d)$. Hence, the number of candidate variables must be reduced  for this method to remain tractable. In this work, we employ the random forest feature importance score to reduce the number of candidate variables.

\section{Results}
\label{sec:results}
\subsection{Algebraic SGS models}
\label{sec:vre_gra}

\emph{A priori} analysis is performed by comparing  SGS stresses $\tau_{ij}^{sgs}$  computed from filtered DNS, with SGS stress modeled by the Vreman (\cref{eq:vreman}) and gradient (\cref{eq:gradient}) models.
The performance of the SGS models is evaluated through the Pearson correlation coefficient, which measures the linear correlation between two variables. 
A Pearson correlation of 1.0 and $-1.0$ corresponds to  perfectly positive and negative linear relationships, respectively, whereas a correlation of 0.0 indicates a negligible linear relationship.

\begin{figure}[htb!]
        \centering
           \begin{subfigure}{0.325\columnwidth}
        \centering
            \includegraphics[width=\columnwidth]{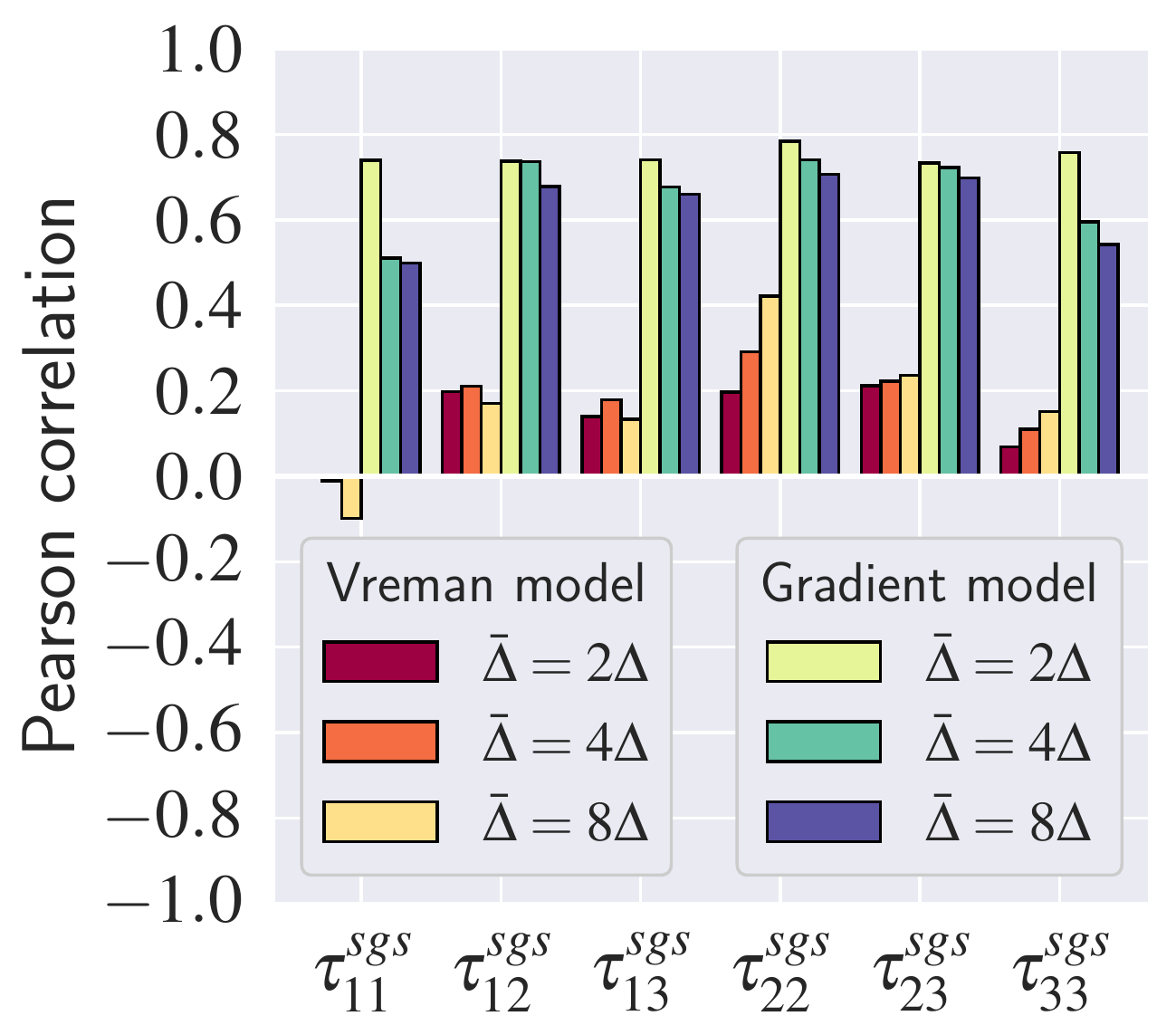}
        \caption{ Inert case. \label{fig:inert_vre_gra}}
    \end{subfigure}
    \begin{subfigure}{0.3\columnwidth}
        \centering
            \includegraphics[width=\columnwidth]{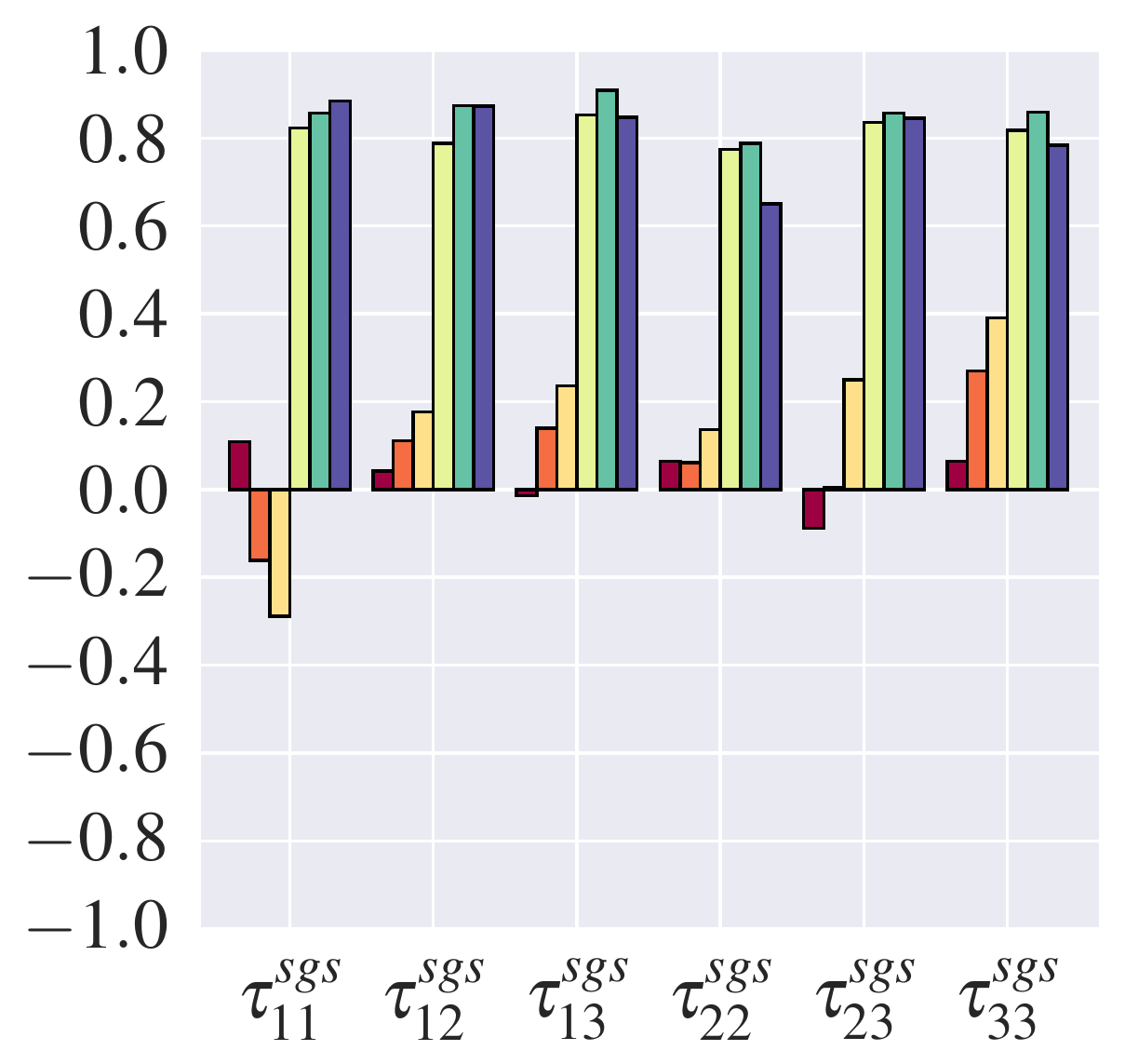}
            \caption{Da $= 780$. \label{fig:reactlarge_vre_gra}}
    \end{subfigure}
        \begin{subfigure}{0.3\columnwidth}
        \centering
            \includegraphics[width=\columnwidth]{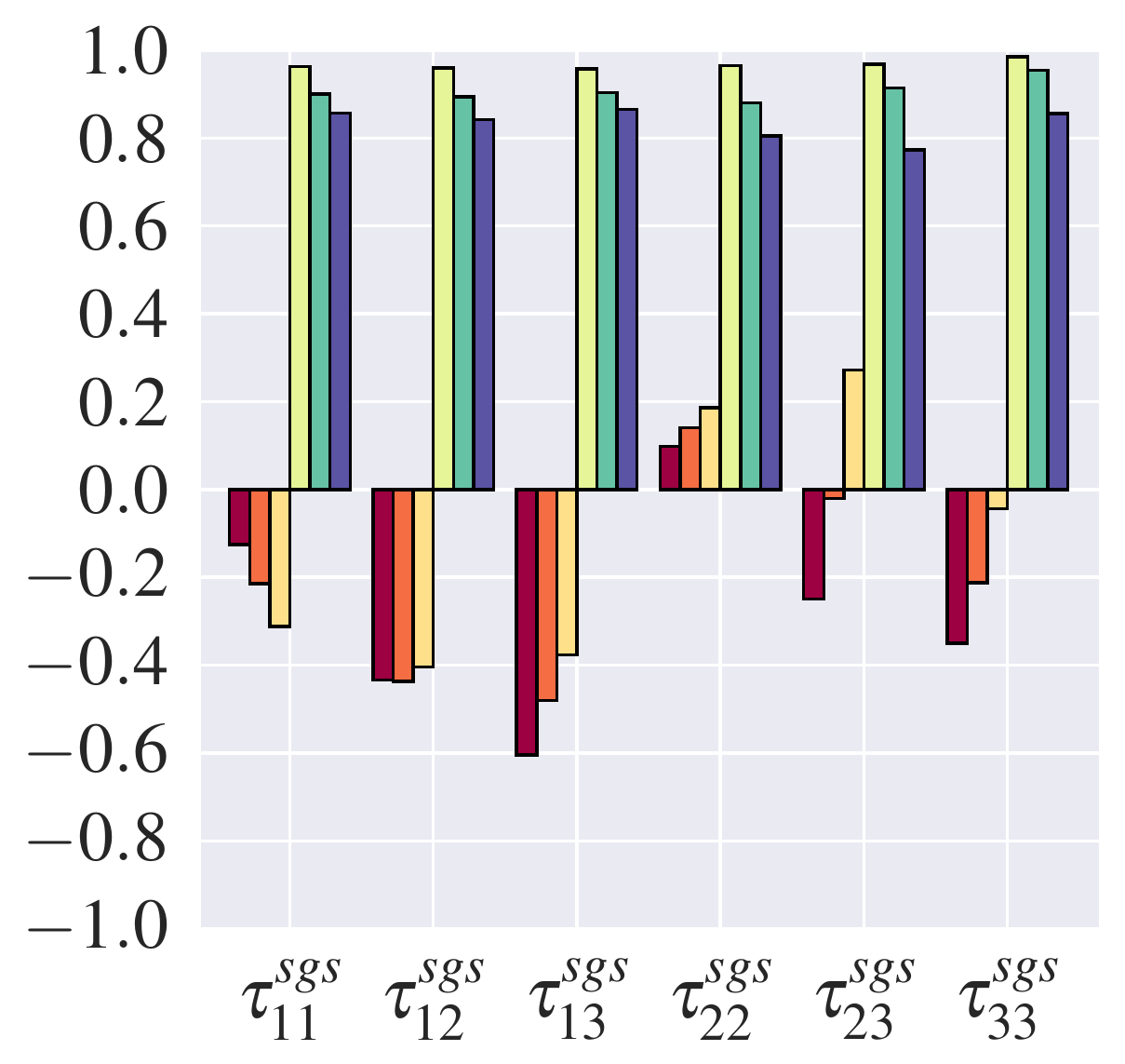}
           \caption{Da $= 10$. \label{fig:reactsmall_vre_gra}}
    \end{subfigure}
    \caption{Pearson correlations between  exact and algebraically modeled SGS stresses for three different filter widths $\ol{\Delta}$. \label{fig:vreman_gradient}}
\end{figure}

\Cref{fig:vreman_gradient} presents the resulting Pearson correlation between  exact and algebraically modeled SGS stresses for three different filter widths $\ol{\Delta}$ for all three DNS cases specified in \cref{tab:init}, at time $t=1.5t_I$. 
For all three cases and filter sizes, negative correlations and weak positive correlations ranging from approximately $-0.6$ to 0.4 are observed for the Vreman model.  
Negative correlations suggest the presence of counter-gradient diffusion, which causes the employment of eddy-viscosity models such as the Vreman model to be ineffective. In all three cases and three filter sizes, strong positive correlations, ranging from 0.5 to 0.95, suggest that the gradient model is highly suitable for modeling SGS stresses in transcritical inert and reacting flows. 

\begin{figure}[htb!]
        \centering
            \begin{subfigure}{0.325\columnwidth}
        \centering
            \includegraphics[width=\columnwidth]{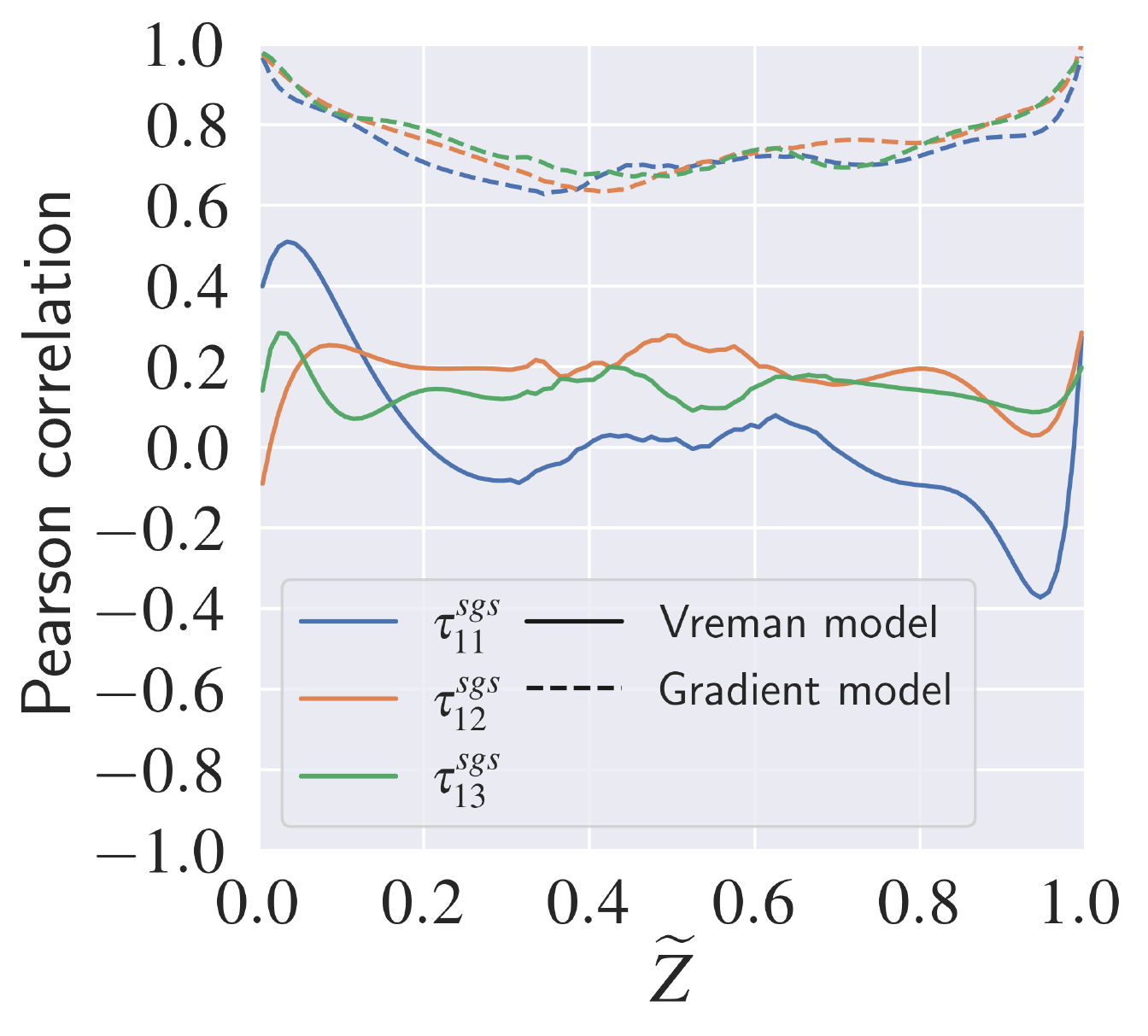}
        \caption{ Inert case. \label{fig:sgs_inert_cond}}
    \end{subfigure}
    \begin{subfigure}{0.3\columnwidth}
        \centering
            \includegraphics[width=\columnwidth]{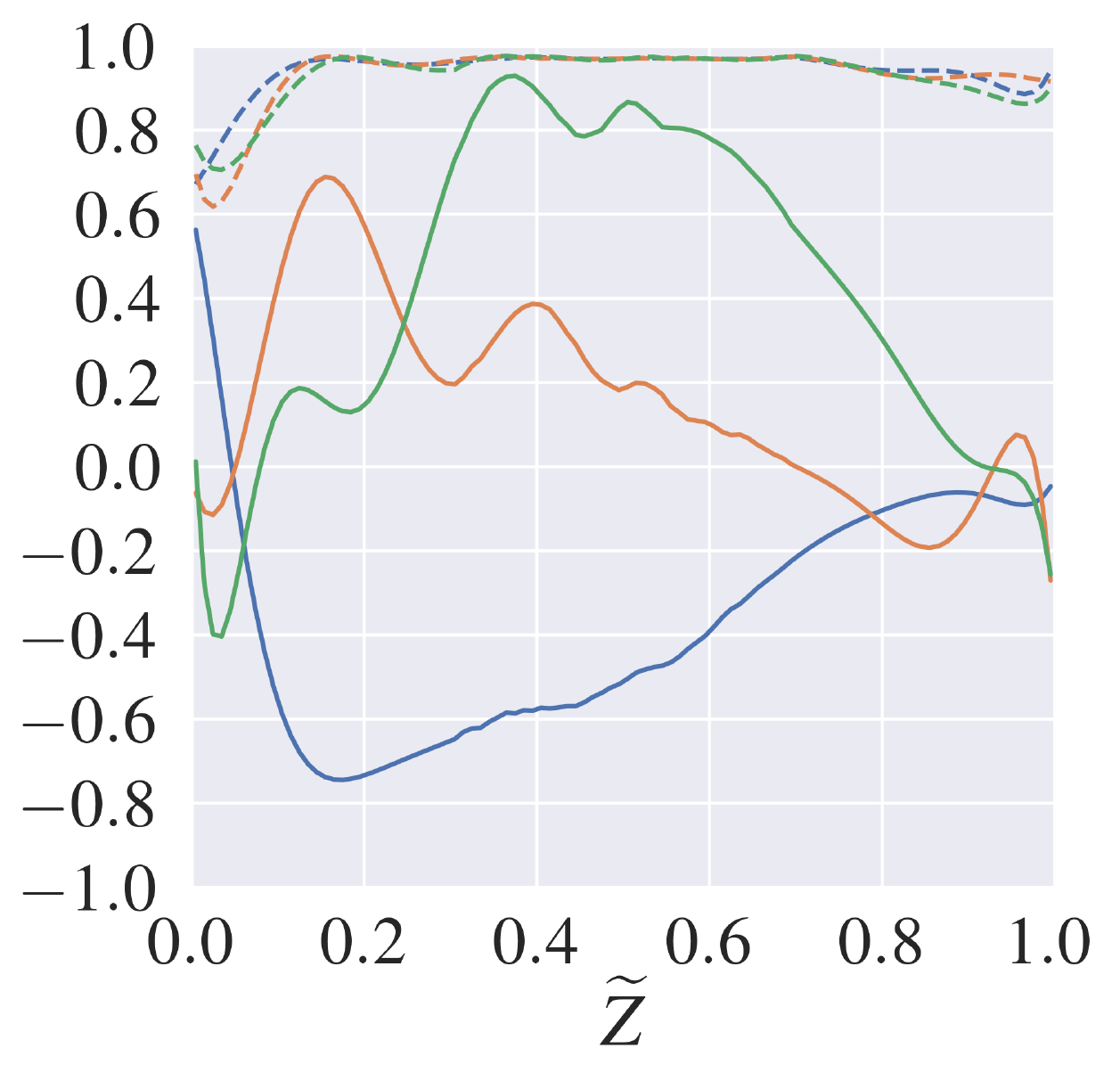}
            \caption{Da $= 780$. \label{fig:sgs_reactlarge_cond}}
    \end{subfigure}
        \begin{subfigure}{0.3\columnwidth}
        \centering
            \includegraphics[width=\columnwidth]{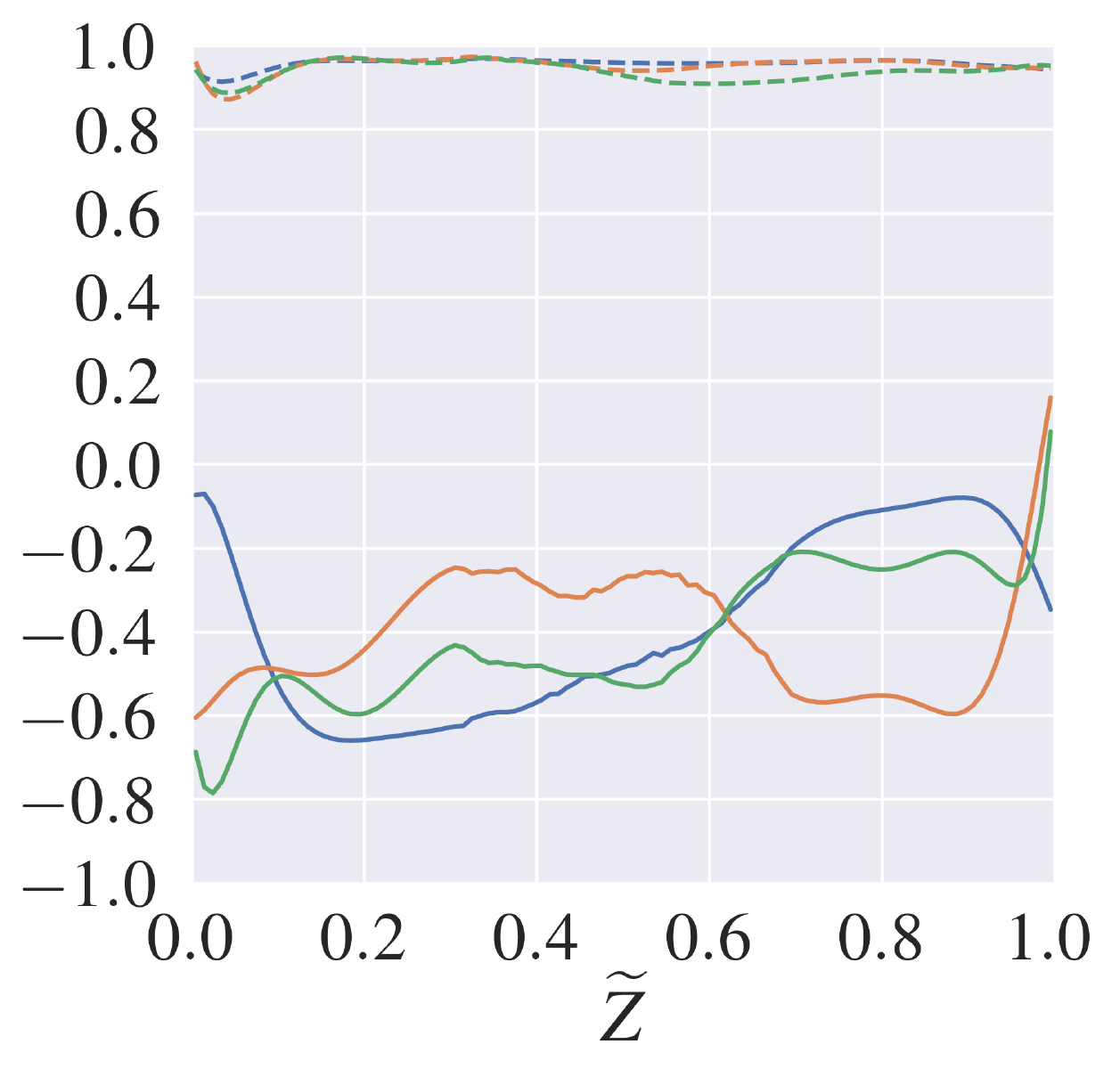}
           \caption{Da $= 10$. \label{fig:sgs_reactsmall_cond}}
    \end{subfigure}
    \caption{Conditional Pearson correlations with respect to mixture fraction $\widetilde{Z}$ between exact and algebraically modeled SGS stresses $\tau^{sgs}_{1i}$ for a single filter width $\ol{\Delta} = 2\Delta$. \label{fig:vreman_gradient_cond}}
\end{figure}

The effectiveness of the Vreman and gradient models are further assessed by examining the conditional Pearson correlation for $\tau^{sgs}_{1i}$  with respect to the mixture fraction $\widetilde{Z}$ at filter size $\ol{\Delta} = 2\Delta$. 
The mixture fraction for the reacting cases have been evaluated using Bilger's definition.
\Cref{fig:sgs_inert_cond} shows that weak correlations ranging from $-0.4$ to 0.5 are observed throughout the inert case.  In both reacting cases in \Cref{fig:sgs_reactlarge_cond,fig:sgs_reactsmall_cond}, the presence of countergradient diffusion is much larger than the inert case, as denoted by the presence of highly negative correlations $(-0.8)$ in the Vreman model.  
In the inert case, the gradient model has the highest correlation of approximately 1.0 in pure methane and pure oxygen, and the lowest correlation of 0.6 when $Z=0.5$. For the case Da $=780$, the gradient model has the lowest correlation (0.7) close to the pure oxygen stream, with the correlation steadily increasing as the mixture approaches stoichiometry ($\widetilde{Z}_{st}=0.2$), after which the correlations remain high (0.85 to 1.0). For the case Da $=10$, the correlations for the gradient model are high (0.8 to 1.0) throughout the entire mixture. 

\begin{figure}[htb!]
        \centering
            \includegraphics[width=0.54\textwidth]{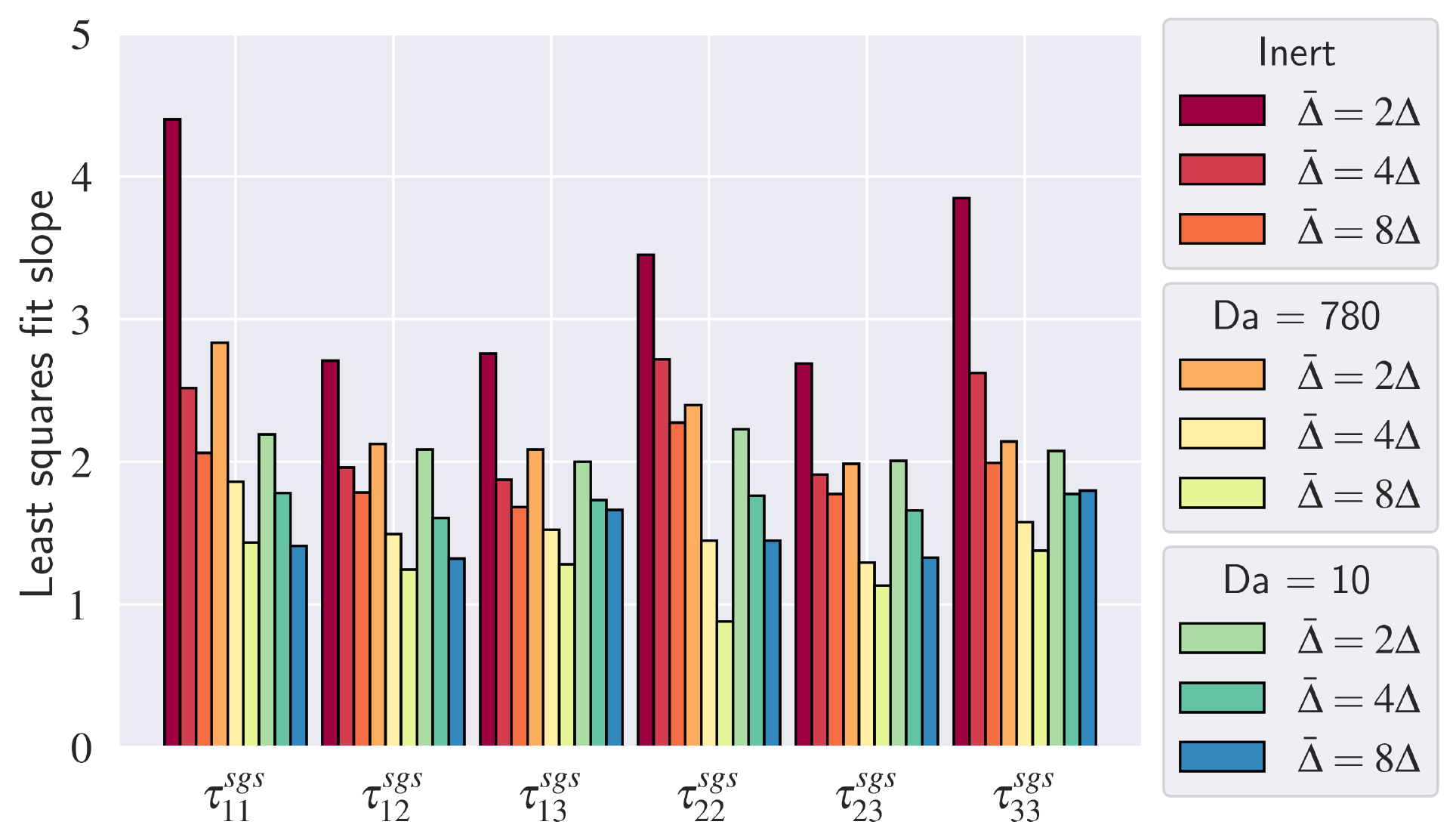}
        \caption{Slopes from a least squares fit of exact and gradient modeled SGS stress for three different filter widths $\ol{\Delta}$. \label{fig:gr_slope}}
\end{figure}

The accuracy of the gradient model in predicting the magnitude of SGS stresses is evaluated by examining the least squares fit between  the exact and modeled SGS stresses. A slope greater than unity indicates underprediction of the modeled SGS stresses, while a slope less than unity indicates overprediction.
\Cref{fig:gr_slope} shows that the slopes from the gradient model range from 1 to 4.5. The average of the slopes is 1.98, which suggests that the gradient model with a constant coefficient should employ $C_g = 1/6$, instead of the typical $C_g =  1/12$. However, since a wide range of coefficients are observed, a dynamic gradient model scheme is likely more suited in \emph{a posteriori} simulations. This is further reinforced by good results from \emph{a posteriori} evaluations of the dynamic gradient model from transcritical inert DNS~\cite{taskinoglu_bellan_2010}.

\subsection{Random forest SGS models}
The \emph{a priori} analysis performed in \cref{sec:vre_gra} is repeated in this section for the SGS stresses modeled by random forest regressors.
\cref{tab:rf} summarizes the random forests employed in this study. 
All random forests are trained with  snapshots at one eddy turnover time $t=t_I$ and tested on the three cases at $t=1.5t_I$. 
Two different sets of feature, or inputs, are employed to train  the random forests. One feature set corresponds to a domain-blind random forest RF\_BLIND, consisting only of velocity, and the first and second spatial derivatives of velocity.
The other set considers Galilean invariant basis functions constructed from strain $\widetilde{S}_{ij}$ and rotation $\widetilde{R}_{ij}$ tensors as features, shown to predict anisotropy well in a previous study~\cite{LING201622}.  These Galilean invariant features are used to train the random forest RF\_INFORM.
In order to investigate the generalizability of random forests in the absence of a vast representative dataset, we evaluate the predictive performance of three additional random forests RF\_INERT,  RF\_DA780, and RF\_DA10, which are trained solely from the inert, Da $= 780$, and  Da $= 10$ cases, respectively.

\begin{table}[ht]
 \centering
 \caption{Random forests employed in this study. \label{tab:rf}}
 \resizebox{\columnwidth}{!}{
\begin{tabular}{|l|c|c|c|c|c|}
\hline 
Random forest &RF\_BLIND & RF\_INFORM & RF\_INERT& RF\_DA780 & RF\_DA10 \\
 \hline  \hline

\multirow{2}{*}{Features (Input)} & \multirow{2}{*}{$\widetilde{u}_i$, $\dfrac{\partial \widetilde{u}_i}{\partial x_j}$, $\dfrac{\partial^2 \widetilde{u}_i}{\partial x_jx_k}$ }& $\widetilde{S}_{ij}$, $\widetilde{S}_{ik}\widetilde{S}_{kj}$, $\widetilde{R}_{ik}\widetilde{R}_{kj}$, &
\multirow{2}{*}{$\widetilde{u}_i$, $\dfrac{\partial \widetilde{u}_i}{\partial x_j}$, $\dfrac{\partial^2 \widetilde{u}_i}{\partial x_jx_k}$ } & \multirow{2}{*}{$\widetilde{u}_i$, $\dfrac{\partial \widetilde{u}_i}{\partial x_j}$, $\dfrac{\partial^2 \widetilde{u}_i}{\partial x_jx_k}$ } & \multirow{2}{*}{$\widetilde{u}_i$, $\dfrac{\partial \widetilde{u}_i}{\partial x_j}$, $\dfrac{\partial^2 \widetilde{u}_i}{\partial x_jx_k}$ }\\ 

& &$\widetilde{S_{ik}}\widetilde{R_{kj}}-\widetilde{R_{ik}}\widetilde{S_{kj}}$& & & \\
\hline
Output & \multicolumn{5}{c|}{$\tau_{ij}^{sgs}$}\\
 \hline
  Training data ($t=t_I$)& All three cases & All three cases &Inert case &Da $=780$ case &Da $=10$ case \\
\hline 
 Testing data ($t=1.5t_I$)&\multicolumn{5}{c|}{All three cases} \\
 \hline
\end{tabular}
}
\end{table}

\begin{figure}[htb!]
        \centering
    \begin{subfigure}{0.44\textwidth}
        \centering
            \includegraphics[width=\textwidth]{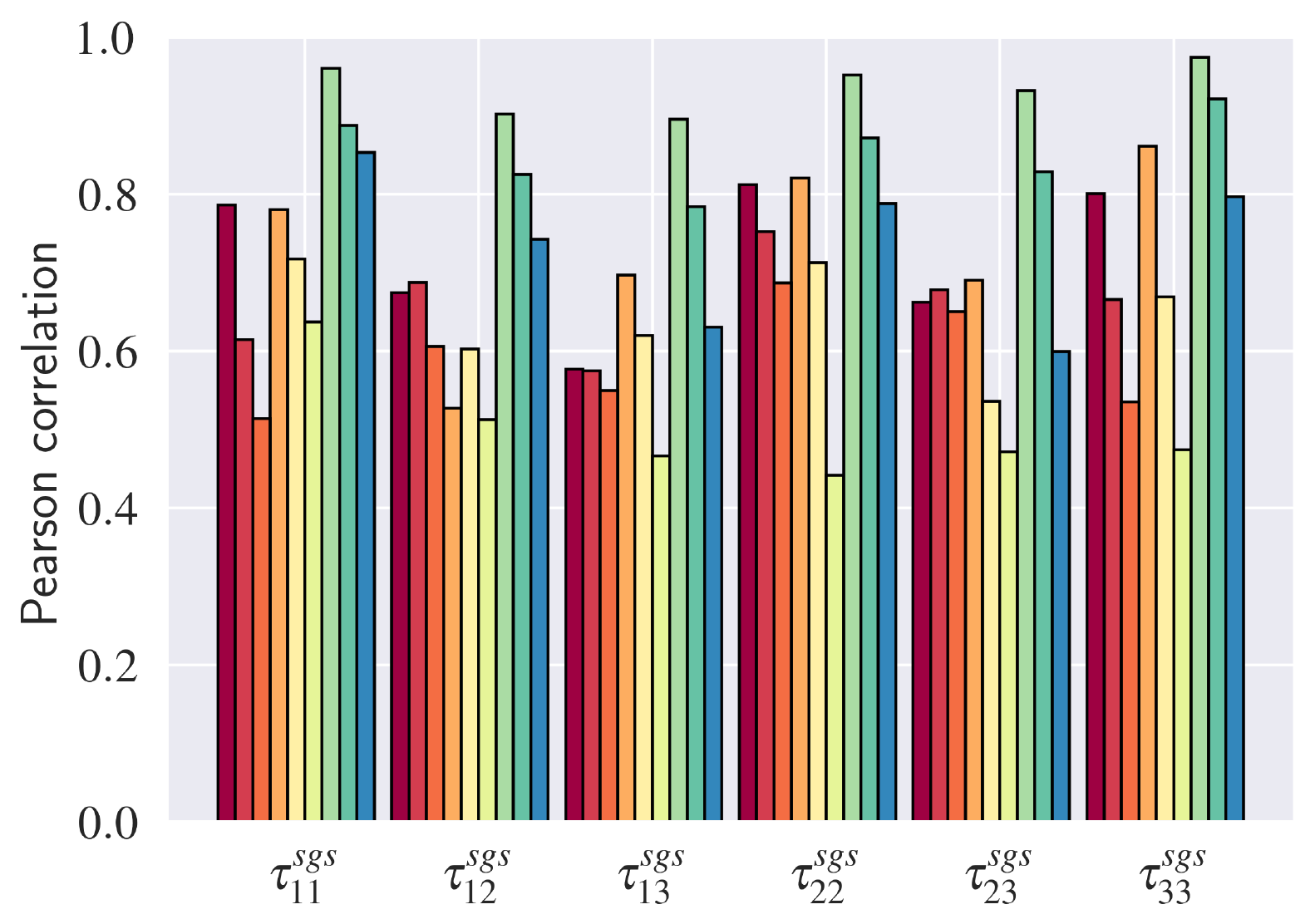}
            \caption{RF\_BLIND. \label{fig:rf_blind_corr}}
    \end{subfigure}
        \begin{subfigure}{0.54\textwidth}
        \centering
            \includegraphics[width=\textwidth]{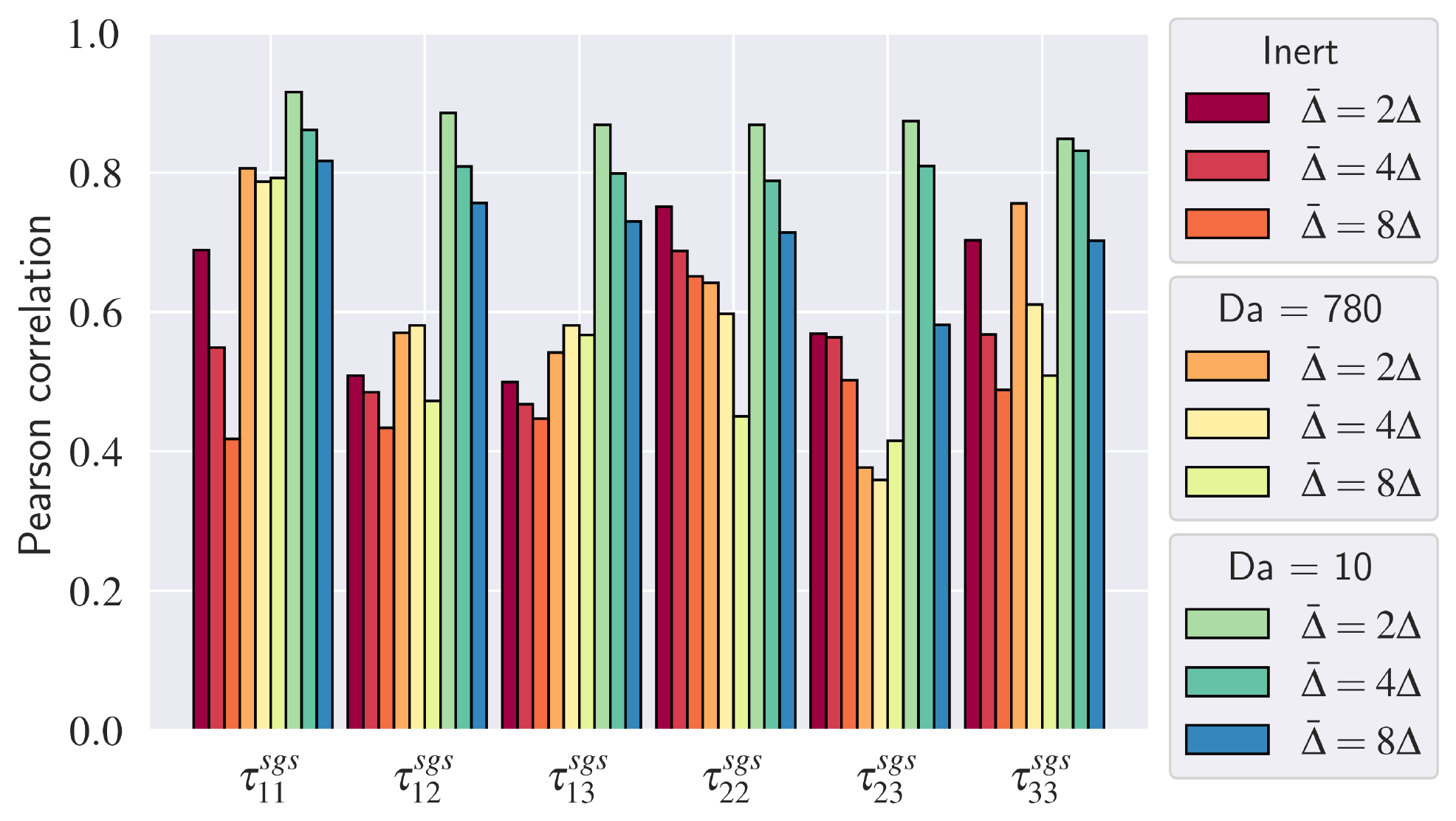}
            \caption{RF\_INFORM. \label{fig:rf_inform_corr}}
    \end{subfigure}
        \caption{Pearson correlation between exact and  random forest modeled SGS stresses for three different filter widths $\ol{\Delta}$.  \label{fig:rf_corr}}

\end{figure}

\Cref{fig:rf_corr} presents the Pearson correlation between exact SGS stresses and the SGS stresses modeled by the random forests RF\_BLIND and RF\_INFORM. \Cref{fig:rf_blind_corr} shows that strong correlations (0.4 to 0.95) are observed when the random forest is trained with an uninformed approach, which is similar to the gradient model and higher than the Vreman model  in \cref{fig:vreman_gradient}. \Cref{fig:rf_inform_corr} demonstrates that the employment  of invariant basis functions as features decreases the range of correlations (0.35 to 0.9) by 0.05. 
This small decrease is likely caused by the additional constraints placed on the random forest when forming a hypothesis space.

\begin{figure}[htb!]
        \centering
            \begin{subfigure}{0.325\columnwidth}
        \centering
            \includegraphics[width=\columnwidth]{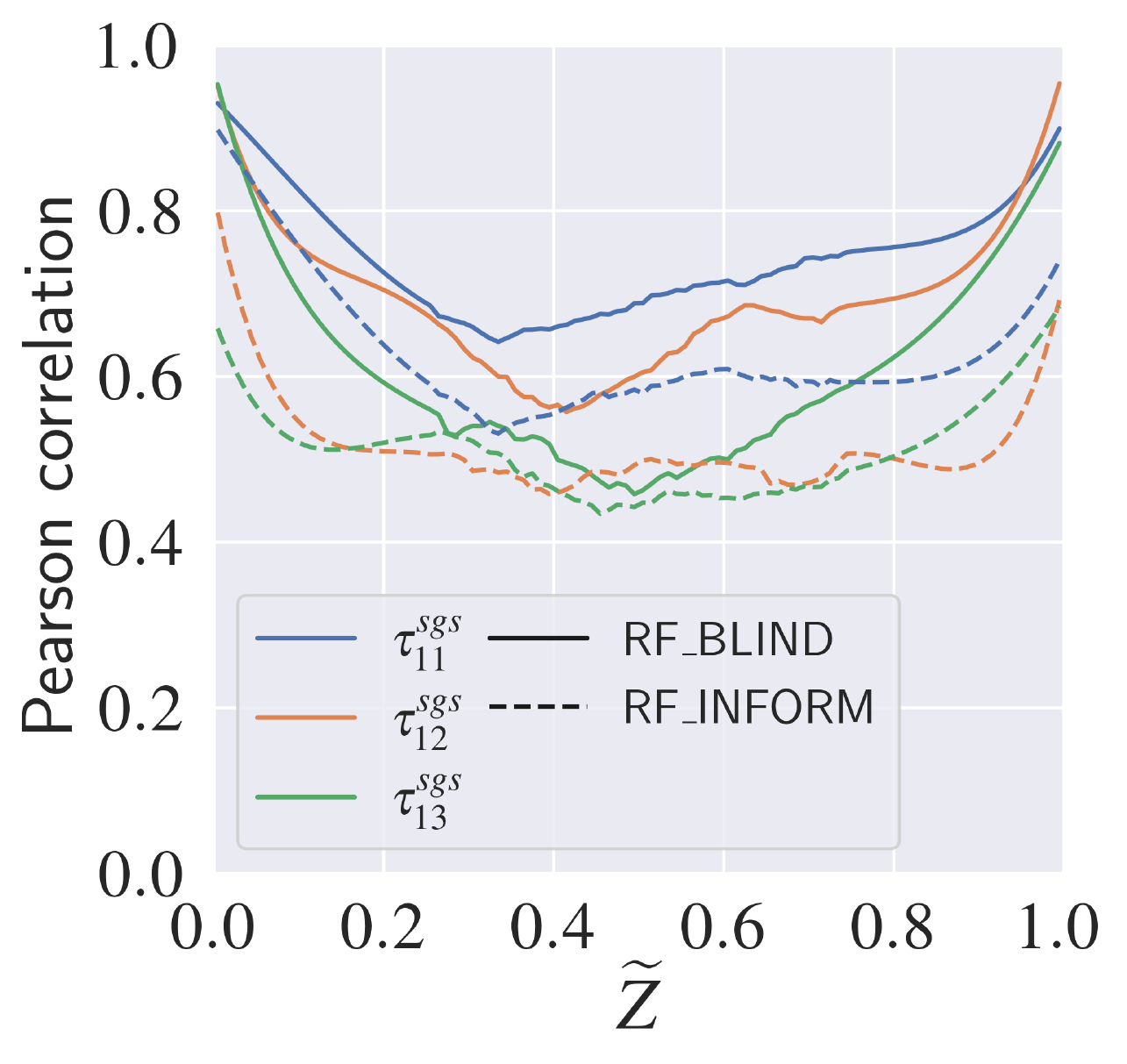}
        \caption{ Inert case. \label{fig:rf_inert_cond}}
    \end{subfigure}
    \begin{subfigure}{0.3\columnwidth}
        \centering
            \includegraphics[width=\columnwidth]{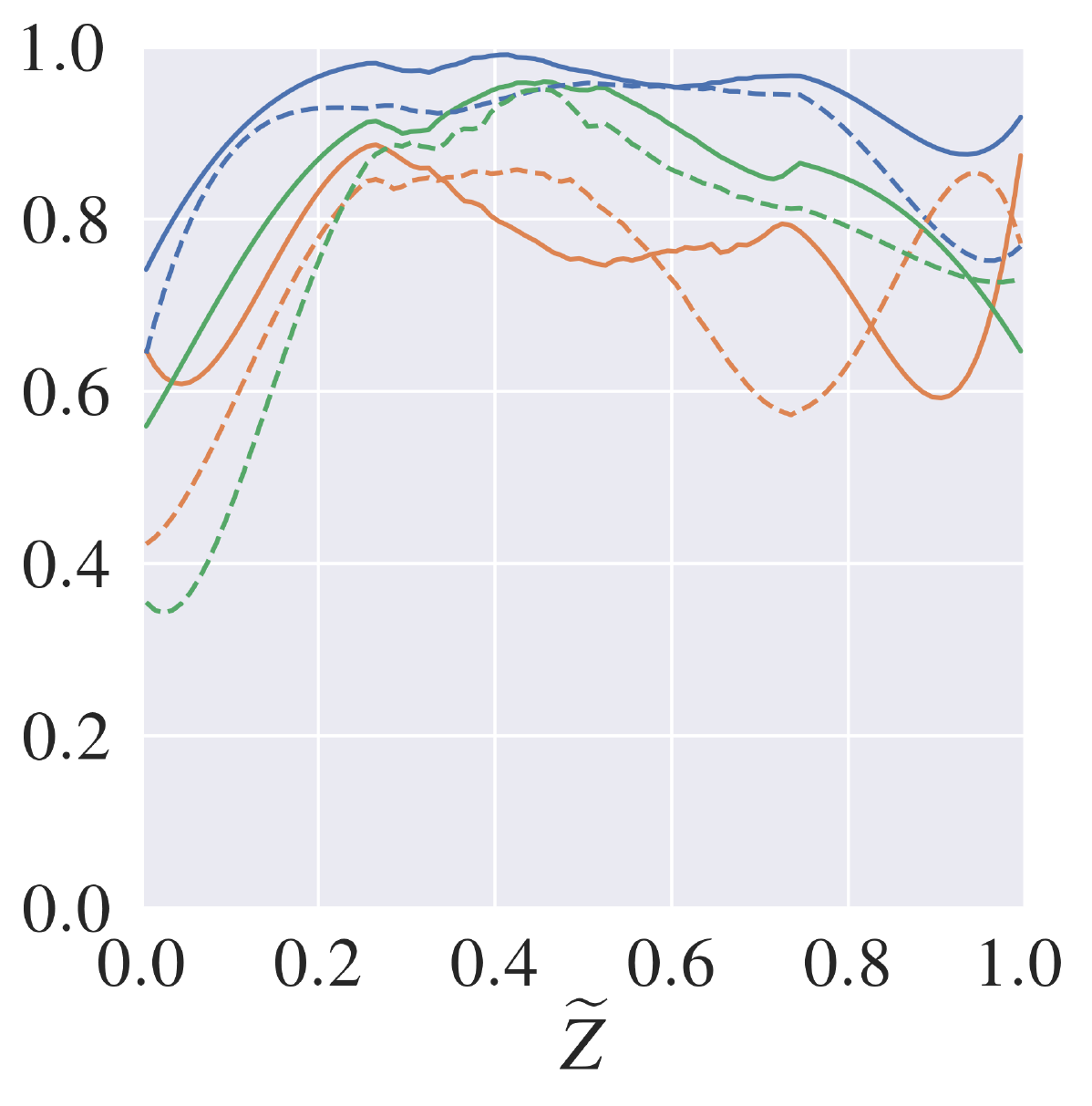}
            \caption{Da $= 780$. \label{fig:rf_reactlarge_cond}}
    \end{subfigure}
        \begin{subfigure}{0.3\columnwidth}
        \centering
            \includegraphics[width=\columnwidth]{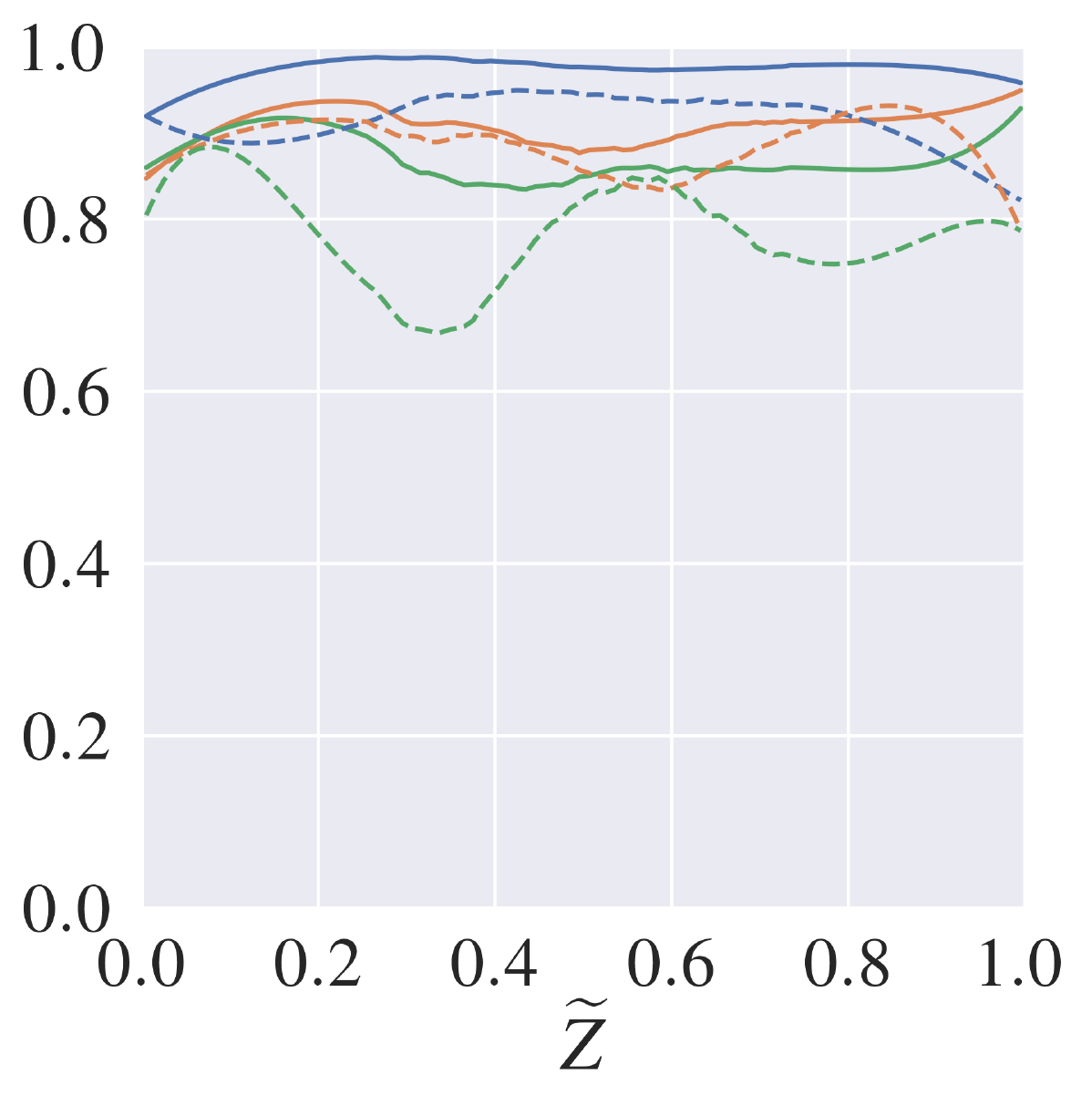}
           \caption{Da $= 10$. \label{fig:rf_reactsmall_cond}}
    \end{subfigure}
    \caption{Conditional Pearson correlations as a function of mixture fraction $\widetilde{Z}$ between exact and random forest modeled SGS stresses $\tau^{sgs}_{1i}$ for a single filter width $\ol{\Delta} = 2\Delta$. \label{fig:rf_cond}}
\end{figure}

\Cref{fig:rf_cond} presents the Pearson correlations between exact and random forest SGS stresses  $\tau^{sgs}_{1i}$  conditioned to  mixture fraction $\widetilde{Z}$ at $\ol{\Delta} = 2\Delta$. 
In the inert case, shown in \cref{fig:rf_inert_cond}, highest correlation from RF\_BLIND of approximately 0.95 is observed in pure methane and pure oxygen, and lowest correlation of 0.5 when $Z=0.5$. 
For the case Da $=780$ in \cref{fig:rf_reactlarge_cond}, RF\_BLIND possesses the lowest correlation (0.7) close to the oxygen stream, with the correlation steadily increasing as the mixture approaches stoichiometric conditions ($\widetilde{Z}_{st}=0.2$), after which the correlations remain high (0.85 to 1.0). 
For the case Da $=10$, shown in \cref{fig:rf_reactsmall_cond}, the correlations for the gradient model are high (0.8 to 1.0) throughout the entire mixture. 
The conditional Pearson correlation produced from RF\_BLIND in all three cases are similar qualitatively and quantitatively to correlations from the gradient model in
\cref{fig:vreman_gradient_cond}.
This suggests that RF\_BLIND has approximated a function similar to the gradient model, even when trained solely on exact SGS stresses and without any prior knowledge of the gradient Model. 
The correlations from  RF\_INFORM share similar  qualitative behaviors as the correlations from RF\_BLIND, but with up to a 0.2 lower values.  

\begin{figure}[htb!]
        \centering
    \begin{subfigure}{0.44\textwidth}
        \centering
            \includegraphics[width=\textwidth]{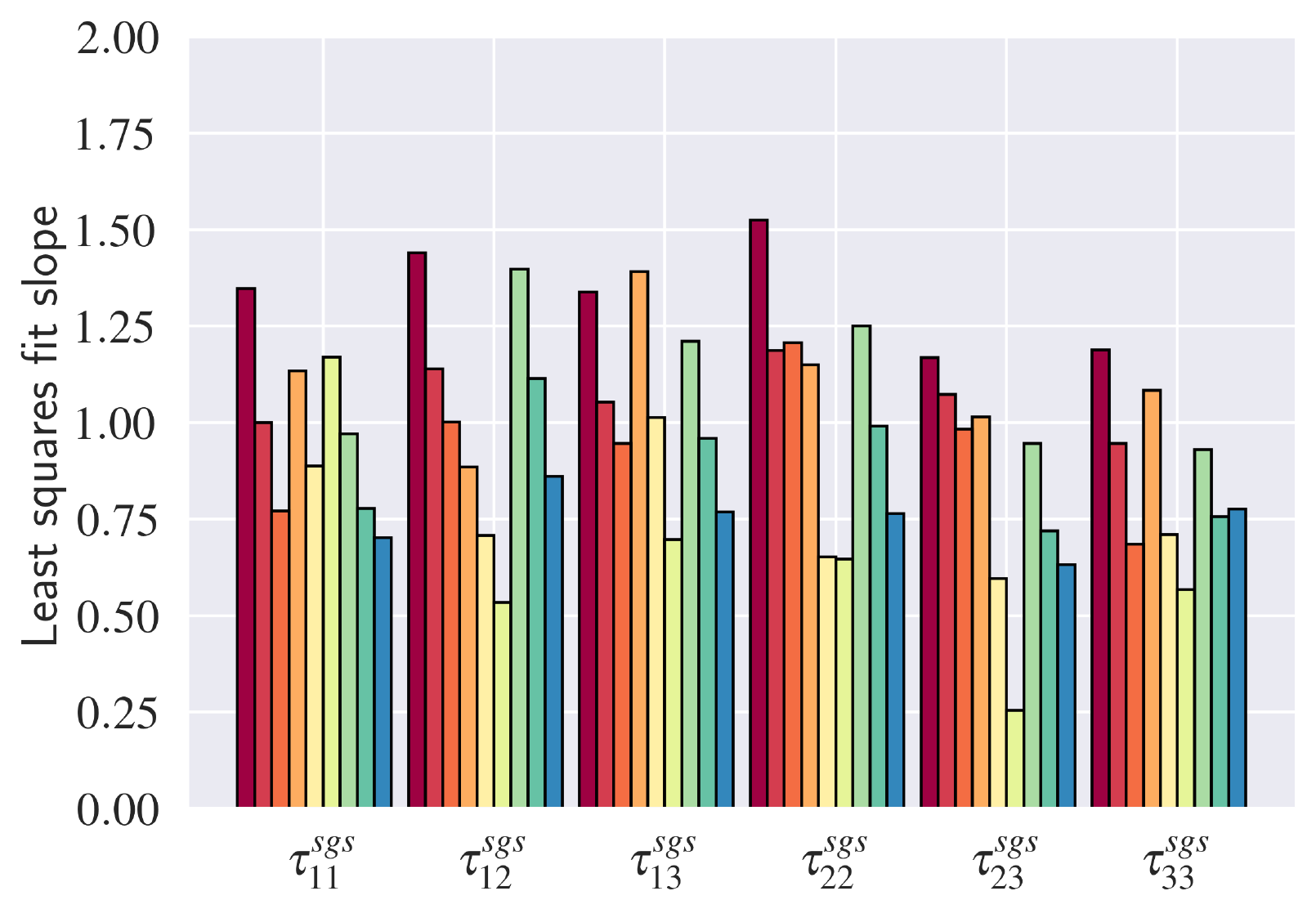}
            \caption{RF\_BLIND. \label{fig:rf_blind_slope}}
    \end{subfigure}
        \begin{subfigure}{0.54\textwidth}
        \centering
            \includegraphics[width=\textwidth]{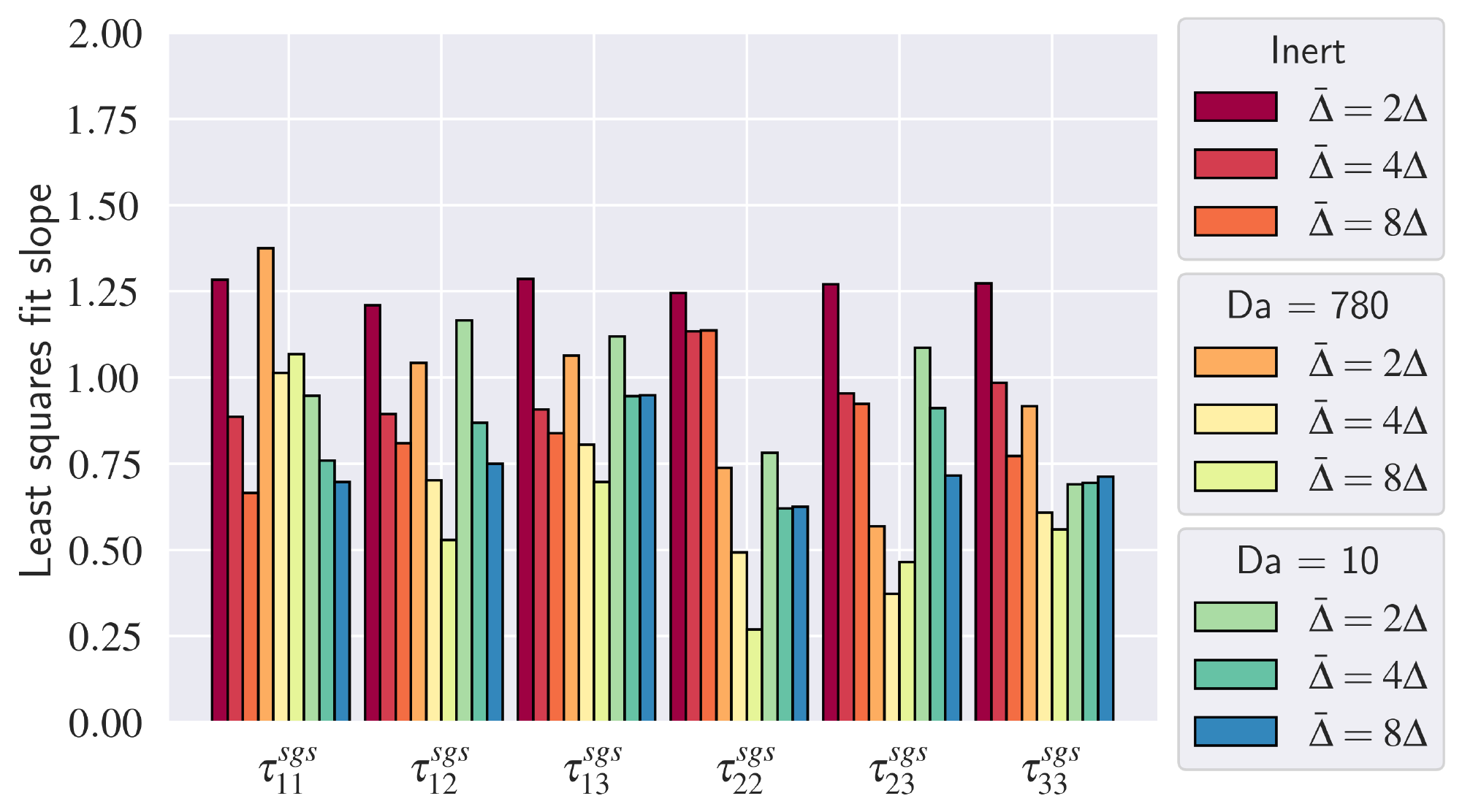}
            \caption{RF\_INFORM. \label{fig:rf_inform_slope}}
    \end{subfigure}
        \caption{Slopes from a least squares fit of exact and random forest modeled SGS stress for three different filter widths $\ol{\Delta}$.  \label{fig:rf_slope}}

\end{figure}

\Cref{fig:rf_slope} presents slopes from a least squares fit between the exact and the random forest SGS stresses. \Cref{fig:rf_blind_slope} shows that the slopes from RF\_BLIND range from 0.25 to 1.6, with an average slope of 0.96, which demonstrates excellent agreement between modeled and exact magnitudes of SGS stresses. The employment of invariant features leads to lower slopes (0.25 to 1.35), with an average slope of 0.867, as presented in \cref{fig:rf_inform_slope}. The use of the invariant feature set not only leads to lower correlations but also to an overprediction in magnitudes of SGS stresses.

\begin{figure*}[htb!]
    \centering
    \includegraphics[width=\textwidth]{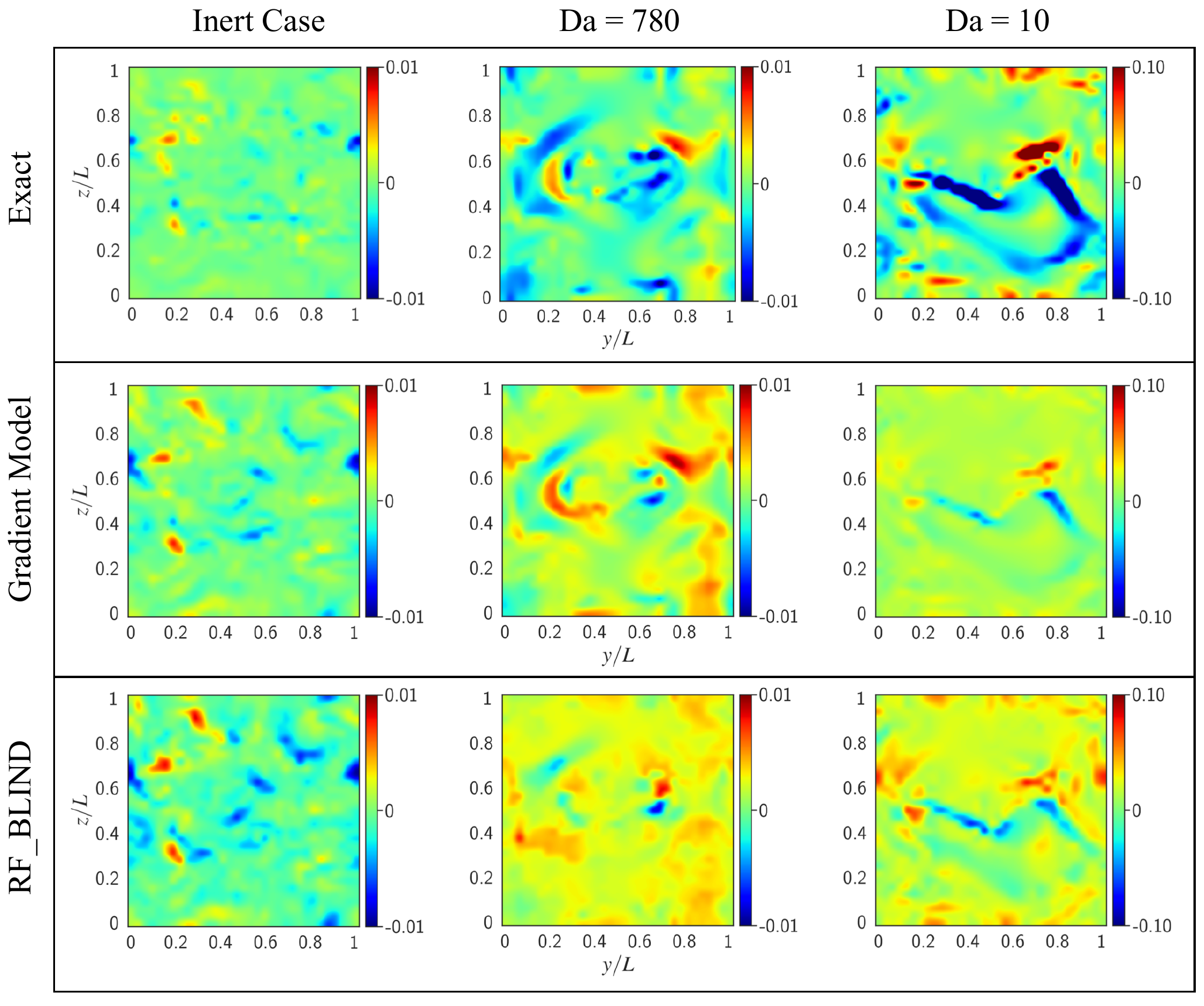}
    \caption{Comparison of exact and modeled SGS stress $\tau^{sgs}_{12}/\ol{\rho}$~[m$^2$s$^{-2}$] at filter width $\ol{\Delta}=4\Delta$ at axial location $x=0$.}
    \label{fig:tau12field}
\end{figure*}

 \Cref{fig:tau12field} compares the exact and modeled SGS stress $\tau^{sgs}_{12}/\ol{\rho}$ at filter width $\ol{\Delta} = 4\Delta$. In the inert case, both SGS stresses from the gradient model and RF\_BLIND are in good agreement with the exact term. For Da~=~780, the gradient model is in  better agreement with the exact term than RF\_BLIND. This is further supported by the difference in Pearson correlation for this particular case shown by the gradient model (0.9) and RF\_BLIND (0.6) in \cref{fig:vreman_gradient,fig:rf_corr}, respectively.  For Da~=~10, RF\_BLIND predicts the magnitude of the SGS stress better than the gradient model, especially near the domain boundaries, which is also observed in the slopes shown by RF\_BLIND (0.9) and  the gradient model (1.5) shown in \cref{fig:gr_slope,fig:rf_slope}.

\begin{figure}[htb!]
        \centering
            \includegraphics[width=0.59\columnwidth]{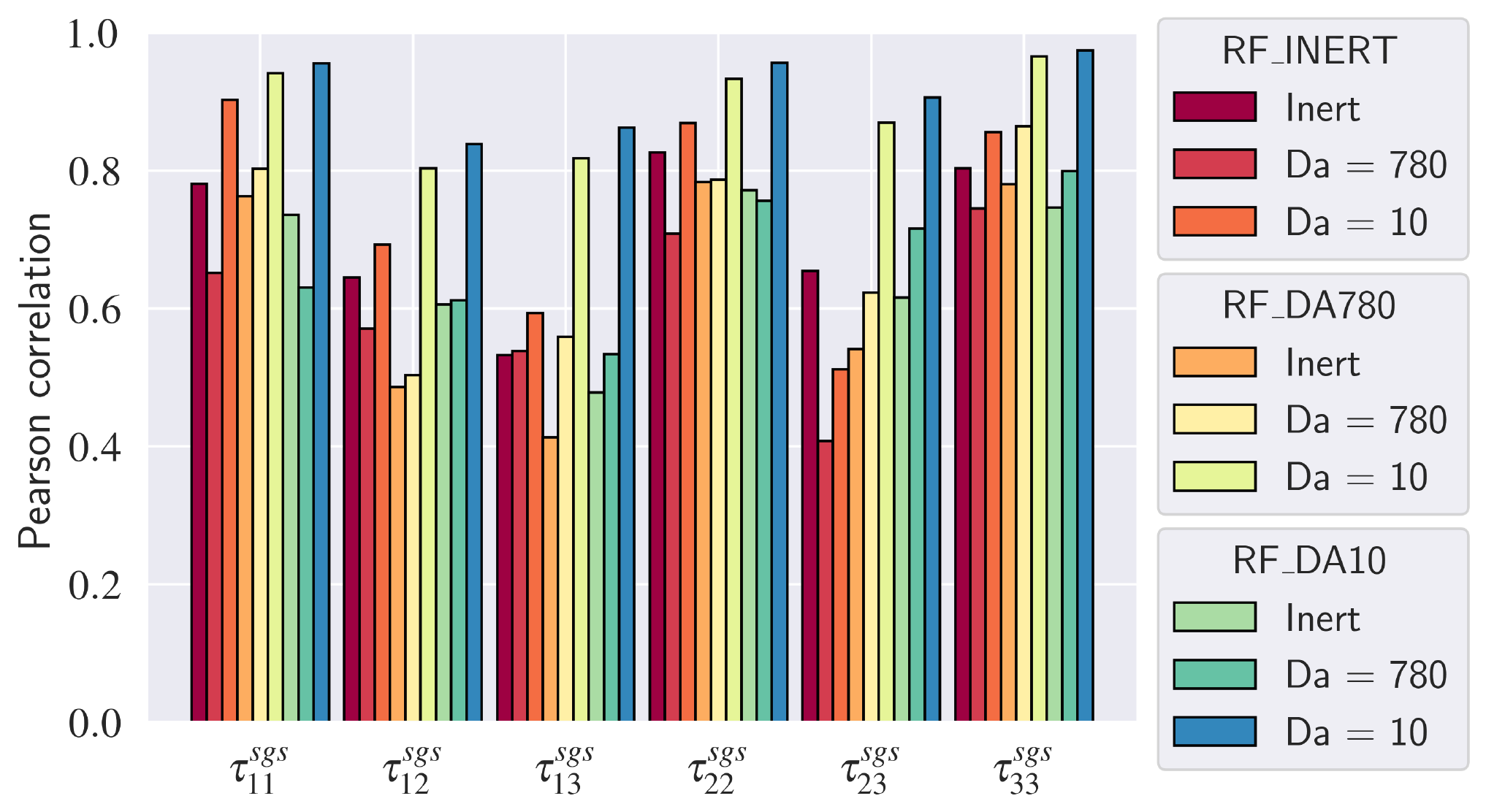}
        \caption{ Pearson correlations between exact and random forest modeled SGS stresses for a single filter width $\ol{\Delta} = 2\Delta$.  \label{fig:rf_general}}
\end{figure}

\Cref{fig:rf_general} presents Pearson correlations from examining the generalizability of random forests in the presence of limited data.
Random forest RF\_INERT demonstrates a similar range of correlations (0.5 to 0.85) to RF\_ALL when tested on the inert case with a filter size  $\ol{\Delta} = 2{\Delta}$. However, lower ranges are observed for RF\_INERT when tested on the cases Da $=780$ (0.4 to 0.75) and Da $=10$ (0.5 to 0.9).  RF\_DA780 also possesses a similar correlation as RF\_BLIND when tested on case Da $=780$ (0.5 to 0.9), but worse correlations when tested on the inert case (0.4 to 0.8) and case Da $=10$ (0.8 to 0.9). Lastly, RF\_DA10 performs similarly to RF\_BLIND when tested on Da $=10$ (0.85 to 0.95) but performs worse when tested on the inert (0.5 to 0.8) and Da $=780$ (0.55 to 0.8) cases. 
These three random forests perform as well as RF\_BLIND on a test set that is represented well by the training set. However, the effectiveness of random forests decreases when modeling on out-of-sample distributions. Nevertheless, these out-of-sample predictions are more accurate than the Vreman model, thus demonstrating a appreciable degree of generalizability.

\subsection{Data-driven discovery of SGS models}
In this section, we examine how the interpretability of random forests can be employed as a tool for model discovery.

\begin{figure*}[htb!]
        \centering            \includegraphics[width=\textwidth]{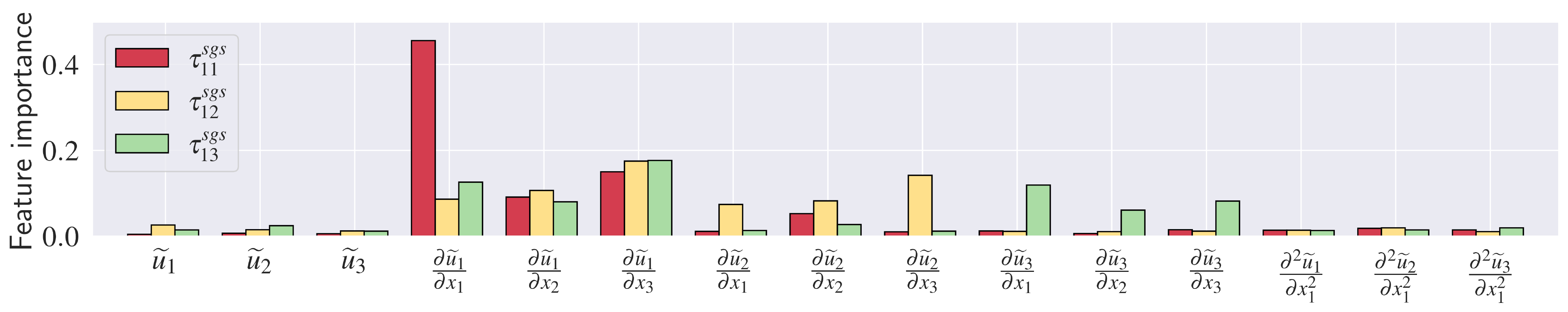}
    \caption{ Fifteen feature importance scores from RF\_BLIND. The other fifteen features, with importance scores less than 0.02, are not shown for brevity. \label{fig:importance}}
\end{figure*}

\Cref{fig:importance} presents  feature importance scores extracted from RF\_BLIND for $\tau^{sgs}_{1i}$. For  all three SGS stresses $\tau^{sgs}_{1i}$ shown, the highest scores are from $\dfrac{\partial \widetilde{u}_1}{\partial x_k}$ and $\dfrac{\partial \widetilde{u}_i}{\partial x_k}$ for three spatial dimensions. 
We employ this observation to formulate a sparse regression problem (see \cref{eq:linearmodel,eq:lsqr,eq:nonlinearcoef}):
\begin{align}
    \frac{\tau^{sgs}_{ij}}{\ol{\rho}u'^2} = f_{ij}\left[G^{d=2} \left( \frac{\ol{\Delta}}{u'}\dfrac{\partial \widetilde{u}_i}{\partial x_k},\frac{\ol{\Delta}}{u'}\dfrac{\partial \widetilde{u}_j}{\partial x_k}\right)\right]
    \label{eq:sgsdiscover}
\end{align}
where the independent variables consist of 2nd-order polynomial functions of the non-dimensionalized selected features. \Cref{eq:sgsdiscover} is non-dimensionalized by density, filter width and initial root-mean-squared velocity  to ensure dimensional consistency in the final model.
This is essential for improving the dimensionality of this sparse regression problem. Since the dimensionality scales with $n^d$ for $n$ number of candidate variables, as discussed in \cref{sec:sparse}, the employment of the feature importance score  for reducing 30 candidate variables to 6 candidate variables results in a  25-fold improvement in dimensionality.

The following equations present the SGS model that resulted from applying sparse regression:
\begin{subequations}
\begin{align}
\tau_{11}^{sgs} & \simeq \ol{\rho} \ol{\Delta}^2 \left( 0.116\dfrac{\partial \widetilde{u}_1}{\partial x_1} \dfrac{\partial \widetilde{u}_1}{\partial x_1} + 0.191 \dfrac{\partial \widetilde{u}_1}{\partial x_2}\dfrac{\partial \widetilde{u}_1}{\partial x_2} + 0.207 \dfrac{\partial \widetilde{u}_1}{\partial x_3}\dfrac{\partial \widetilde{u}_1}{\partial x_3} \right) \label{eq:tau11}\\
\tau_{12}^{sgs} & \simeq \ol{\rho} \ol{\Delta}^2 \left( 0.113\dfrac{\partial \widetilde{u}_1}{\partial x_1}\dfrac{\partial \widetilde{u}_2}{\partial x_1} +
0.102\dfrac{\partial \widetilde{u}_1}{\partial x_2}\dfrac{\partial \widetilde{u}_2}{\partial x_2} + 0.134 \dfrac{\partial \widetilde{u}_1}{\partial x_3}\dfrac{\partial \widetilde{u}_2}{\partial x_3} \right)\label{eq:tau12}\\
\tau_{13}^{sgs} & \simeq \ol{\rho} \ol{\Delta}^2 \left( 0.119\dfrac{\partial \widetilde{u}_1}{\partial x_1}\dfrac{\partial \widetilde{u}_3}{\partial x_1} + 0.117\dfrac{\partial \widetilde{u}_1}{\partial x_2}\dfrac{\partial \widetilde{u}_3}{\partial x_2} + 0.109 \dfrac{\partial \widetilde{u}_1}{\partial x_3}\dfrac{\partial \widetilde{u}_3}{\partial x_3} \right)\label{eq:tau13}\\
\tau_{22}^{sgs} & \simeq \ol{\rho} \ol{\Delta}^2 \left( 0.215\dfrac{\partial \widetilde{u}_2}{\partial x_1}\dfrac{\partial \widetilde{u}_2}{\partial x_1} + 0.135\dfrac{\partial \widetilde{u}_2}{\partial x_2}\dfrac{\partial \widetilde{u}_2}{\partial x_2} + 0.164 \dfrac{\partial \widetilde{u}_2}{\partial x_3}\dfrac{\partial \widetilde{u}_2}{\partial x_3} \right)\label{eq:tau22}\\
\tau_{23}^{sgs} & \simeq \ol{\rho} \ol{\Delta}^2 \left( 0.123\dfrac{\partial \widetilde{u}_2}{\partial x_1}\dfrac{\partial \widetilde{u}_3}{\partial x_1} + 0.116\dfrac{\partial \widetilde{u}_2}{\partial x_2}\dfrac{\partial \widetilde{u}_3}{\partial x_2} + 0.134 \dfrac{\partial \widetilde{u}_2}{\partial x_3}\dfrac{\partial \widetilde{u}_3}{\partial x_3} \right)\label{eq:tau23}\\
\tau_{33}^{sgs} & \simeq \ol{\rho} \ol{\Delta}^2 \left( 0.251\dfrac{\partial \widetilde{u}_3}{\partial x_1}\dfrac{\partial \widetilde{u}_3}{\partial x_1} + 0.177\dfrac{\partial \widetilde{u}_3}{\partial x_2}\dfrac{\partial \widetilde{u}_3}{\partial x_2} + 0.124 \dfrac{\partial \widetilde{u}_3}{\partial x_3}\dfrac{\partial \widetilde{u}_3}{\partial x_3} \right) \label{eq:tau33}
\end{align}
\end{subequations}
The resulting model can be rewritten as:
\begin{align}
    \tau_{ij}^{sgs} & \simeq \ol{\rho} \ol{\Delta}^2 \left( C_1\dfrac{\partial \widetilde{u}_i}{\partial x_1} \dfrac{\partial \widetilde{u}_j}{\partial x_1} + C_2 \dfrac{\partial \widetilde{u}_i}{\partial x_2}\dfrac{\partial \widetilde{u}_j}{\partial x_2} + C_3 \dfrac{\partial \widetilde{u}_i}{\partial x_3}\dfrac{\partial \widetilde{u}_j}{\partial x_3} \right) \label{eq:tauij}
\end{align}
where the resulting model coefficients $C_{\{1,2,3\}}$ range from 0.102 to 0.251. 
\Cref{eq:tauij} is  similar in form to the gradient model (\cref{eq:gradient}), but possesses three model coefficients instead of one.
By observing that $C_{\{1,2,3\}}$ are of the same order of magnitudes, and collapsing the three coefficients by evaluating  the average model coefficients, we recover the gradient model:
\begin{align}
    \tau_{ij}^{sgs} & \simeq \ol{\rho}  C_{x} \ol{\Delta}^2 \dfrac{\partial \widetilde{u}_i}{\partial x_k}\dfrac{\partial \widetilde{u}_j}{\partial x_k}
\end{align}
where the model coefficient $C_{x} = 0.147$ is similar in value to the suggested model coefficient of 0.167 from \cref{sec:vre_gra}.
This result  demonstrates that the employment of sparse regression, in conjunction with random forest feature importance can be employed to discover an algebraic expression, similar to the effective gradient model, for modeling subgrid scale stresses in transcritical flows.

Since the present method relies on the random forest feature importance score, a statistical test must be employed to test for the effects of significant correlation amongst the features. If multiple features in the modeling basis are significantly correlated, they act as exchangeable surrogates for each other during the calculation of feature importance scores (this is similar to the phenomenon of multicollinearity in classical statistics~\cite{kumar1975multicollinearity}). Under such conditions, metrics such as the MDI are susceptible to correlation bias, and can generate erroneous importance scores~\citep{altmann2010permutation, tolocsi2011classification}. As a note, almost all algorithms for estimating feature importance, including SHAP (Shapley additive explanations)~\cite{lundberg2017unified} exhibit such correlation bias. As an alternative, Principal Component Analysis may be utilized to engender orthogonal bases for new features that are independent. However, these derived features are often difficult to ascribe physical meanings to, obfuscating their utility toward interpretability.  

\begin{figure}[htb!]
        \centering
            \includegraphics[width=0.9\columnwidth]{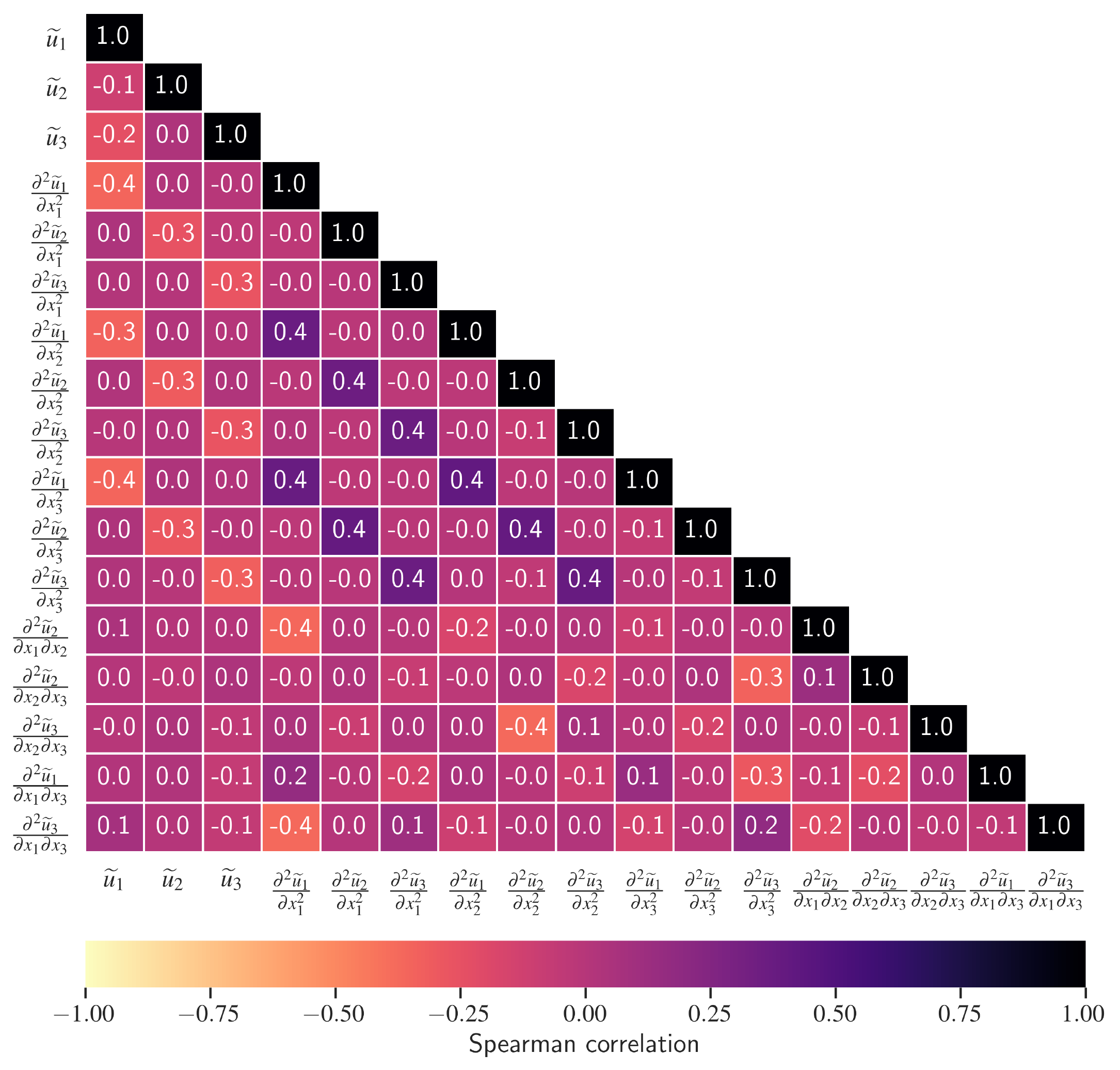}
        \caption{Spearman correlation matrix for selected features from RF\_BLIND.  Features with correlations less than 0.2 are not shown for brevity. \label{fig:spearman}}
\end{figure}

Multicollinearity is likely not an issue for RF\_BLIND since we know from finite differencing schemes that the spatial derivative of a quantity in a cell cannot be evaluated locally, but must be evaluated from neighboring cells. Nevertheless, we utilize the Spearman correlation as a statistical test for evaluating the correlation amongst the features in the modeling basis. While the Pearson correlation is a statistical tool used for evaluating linear relationships, the Spearman correlation evaluates the monotonicity of variables in both linear and non-linear functions, i.e., whether the increasing or decreasing trend is being preserved. Spearman correlations of 1 and $-1$ correspond to a perfect monotonic relationship, while 0 corresponds to a negligible monotonic relationship.
\Cref{fig:spearman} shows that Spearman correlations between different features from RF\_BLIND are weak (between $-0.4$ and 0.4), which indicates weak correlation amongst the features.

\section{Conclusions} \label{sec:conc}

DNS of inert and reacting transcritical LOX/GCH4 non-premixed mixtures under decaying turbulence were performed. 
 Pressure and temperature were chosen to correspond to conditions in rocket combustors. 
\textit{A priori} analysis was  conducted by comparing exact subgrid-scale stresses from Favre-filtered DNS data with algebraic and data-driven SGS models. 

\textit{A priori} analysis showed that the SGS stresses evaluated by Vreman SGS model correlated poorly with the corresponding exact terms, likely caused by the presence of countergradient diffusion. In contrast, good correlations are seen from the gradient SGS model. Results demonstrated a wide range of magnitude errors in the gradient model, which suggests that a dynamic gradient model approach is suited in \emph{a posteriori} simulations. 
Random forests demonstrated high correlations when trained on datasets which are representative of the test sets, with reasonable predictions for the magnitude of subgrid-scale stresses.
However, correlations were shown to decrease significantly when tested out-of-sample.

Sparse regression was performed to discover an algebraic expression for SGS stresses from non-linear transformations of velocity and its derivatives. The interpretability of random forests was demonstrated to reduce the dimensionality of the sparse regression problem by 25 times, by employing the feature importance score for variable selection. The derived algebraic expression was shown to be similar to the gradient model.  

Results demonstrate that random forests can perform as effectively as suitable algebraic models when modeling subgrid stresses, if trained on a sufficiently representative database. However, in the absence of such a database, this good performance is not replicated. Nevertheless, the employment of random forests can provide insight into the discovery of subgrid-scale models through the feature importance score.
Additional work on interpretable deep learning, which take advantage of autoencoders, could be an exciting direction for future work. 
In addition, the present study should be complemented with an \textit{a posteriori} study, and extended to other SGS closure terms that form chemical source terms and non-linear state equations, to generate further insight.

\section*{Acknowledgments}
The authors gratefully acknowledge financial support from NASA with award No. 80NSS C18C0207, the Department of Energy, National Nuclear Security Administration under award No. DE-NA0003968, and the German Research Foundation (Deutsche
Forschungsgemeinschaft -- DFG) in the framework of
the Sonderforschungsbereich Transregio 40.
Resources supporting this work are provided by the National Energy Research Scientific Computing Center, a U.S. Department of Energy Office of Science User Facility operated under contract No. DE-AC02-05CH11231.

\bibliography{paperDACombustion.V01.bib} 
\bibliographystyle{elsarticle-num-PROCI_titlemod.bst}


\end{document}